\documentclass [12pt]{article}
\usepackage[cp1251]{inputenc}
\usepackage[english]{babel}
\usepackage{amssymb}
\usepackage{epsfig}
\usepackage{float}
\usepackage{slashed}
\newcommand{\be}{\begin{equation}}
\newcommand{\ee}{\end{equation}}
\newcommand{\bea}{\begin{eqnarray}}
\newcommand{\eea}{\end{eqnarray}}

\def\p{\partial}
\def\pslash{\p\raise.3ex \hbox{\kern-.5em /}}
\def\delslash{\nabla\raise.3ex \hbox{\kern-.7em /}}

\begin{document}

\begin{flushright}\bf{In blessed memory\\
of Vladimir Georgievich Kadyshevsky}\end{flushright}
\begin{center}
\huge{\textbf{An Algebraic PT-Symmetric Quantum Theory with a
Maximal Mass}}\footnote{ ISSN 1063-7796, Physics of Particles and
Nuclei, 2016, Vol. 47, No. 2, pp. 135–156. © Pleiades Publishing,
Ltd., 2016. Original Russian Text © V.N. Rodionov, G.A. Kravtsova,
2016, published in Fizika Elementarnykh Chastits i Atomnogo Yadra,
2016, Vol. 47, No. 2.  DOI: 10.1134/S1063779616020052}
\end{center}

\begin{center} \Large{\textbf {V. N. Rodionov${}^a$ and G. A.
Kravtsova}${}^b$}
\end{center}
\begin{center}
${}^a$ Plekhanov Russian University of Economics, Faculty of
Mathematical Economics and Informatics, Stremyannyi per. 36,
Moscow, 117997 Russia e-mail: rodyvn@mail.ru
\end{center}
\begin{center}
${}^b$ Moscow State University, Faculty of Physics, Vorob’evy gory
1, Moscow, 119991 Russia e-mail: gakr@chtc.ru
\end{center}

                      \vskip 0.5cm \begin{flushright}
                       \it{“The stone which the builders
                      rejected\\ has become the head of the corner” (Psalm
                       17:22-23).}\end{flushright}
\begin{center}

\abstract{ In this paper we draw attention to the fact that the
studies by V. G. Kadyshevsky devoted to the creation of the  which
\emph{\emph{to the geometric quantum field theory with a
fundamental mass}} containing non-Hermitian mass extensions. It is
important that these ideas recently received a powerful
development in the form of construction of the non-Hermitian
algebraic approach. The central point of these theories is the
construction of new scalar products in which the average values of
non-Hermitian Hamiltonians are valid. Among numerous works on this
subject may be to allocate as purely mathematical and containing a
discussion of experimental results. In this regard, we consider as
the development of algebraic relativistic pseudo-Hermitian quantum
theory with a maximal mass and experimentally significant
investigations are discussed}

\end{center}

\section{ Introduction}

This paper was planned as a joint study with V. G. Kadyshevsky.
Fate decreed otherwise. A great person and an outstanding
scientist, Academician of the Russian Academy of Sciences Vladimir
Georgievich Kadyshevsky passed away. We were fortunate to begin
studies in the field of the theory which is rightfully considered
his creation. This theory was called the \emph{quantum field
theory (QFT) with a fundamental mass}, i.e., a modified QFT whose
basis includes, along with common postulates of the quantum
theory, a new fundamental principle stating that the mass spectrum
of elementary particles should be limited from above, $m\leq M$.

At present, it is assumed that elementary particles are those
particles whose properties and interactions can be adequately
described in terms of local fields. The following question can
also be formulated in these terms: should the mass of elementary
particles be limited from above? Namely, to what values of the
particle mass m is the concept of a local field applicable for
describing this particle?

Vladimir Georgievich wrote in this relation: "Formally, the
standard QFT remains logically faultless, even if objects whose
masses are about the mass of a car participate in an elementary
interaction. Such a far extrapolation of the local field theory
toward macroscopic mass values seems a pathology and has hardly
anything in common with the needs of elementary particle physics.
We repeat, however, that the modern QFT does not forbid such a
meaningless extrapolation. May this be a fundamental defect of the
theory, its "Achilles heel"?" [1].

Should a mass of elementary particles be limited from above? Many
scientists state that they do not "believe" in such a constraint.
This problem, however, is not a question of belief. A scientific
question can receive the final answer in experiment only. Until
now, special experiments on the search of particles with the
maximal mass have not been formulated. It is only known that at
present, the most massive particle in the Standard Model (SM) is
the top quark whose mass exceeds the electron mass by
approximately a factor of 300000. It is clear that the search of
direct experiments on detection of "maximons" is limited by the
capabilities of super-high power accelerator facilities. Detailed
investigation of models with a maximal mass may open quite new
unique possibilities for detecting the consequences of this
constraint. We speak of taking into account various external
actions that make it possible to find effects determined by the
limited character of the mass spectrum of elementary particles.
This can be exemplified by the investigation of the influence of
high-intensity magnetic fields on such processes; taking into
account interaction with such fields may result in observability
of a number of effects. In particular, one of the possible
consequences of the limited character of the mass spectrum is the
appearance in the theory of the so-called "exotic" particles whose
existence was predicted by Kadyshevsky in the framework of the
geometric approach [2]. The properties of these particles
cardinally differ from those of their ordinary partners. It turned
out, however, that the appearance of "exotic" particles in the
theory is not a prerogative of the geometric approach. Indeed, the
development of the pseudo-Hermitian algebraic PT-symmetric theory
showed that these particles emerge as a consequence of the
\emph{limited character of the mass spectrum of elementary
particles itself}. Thus, experiments on the search of "exotic"
particles may result in detection of existence of the limiting
mass. This approach becomes realistic due to the calculation of
the energy spectrum of a neutral fermion possessing an anomalous
magnetic moment in the theory with a maximal mass [3, 4]. Thus,
further development of the theory founded by Kadyshevsky can and
should yield proposals on formulation of such experiments in the
near future.

\section{The geometric theory with a limited mass:
scalar and fermion sectors}

The idea of a limited character of the mass of elementary
particles was put forward in 1965 by M.A. Markov. This constraint
was connected with the "Planck mass" $m_p = \sqrt{\hbar c / G}
\sim {10}^{19}$ GeV, where $G$ is the gravitational constant,
$\hbar$ is the Planck constant, and $c$ is the speed of light, and
was written as follows [5]:
\begin{equation}\label{mpl}m\le
m_{Planck} = 10^{19} {\it GeV}. \end{equation} Particles with the
limiting mass $ m = m_{Planck} $ were called by the author
"maximons". They occupied a special place among elementary
particles; in particular, in Markov's scenario of the early
Universe, maximons played an important role [6]. Original
condition (1), however, was purely phenomenological, and actually
did not participate in the construction of the theory. A new
radical approach to introduction of the limited mass spectrum in
the theory was proposed in late 1970 by Kadyshevsky [2]. In this
approach, Markov's idea on existence of a maximal mass of
particles was taken as a new fundamental physical principle of the
quantum field theory. In the proposed theory, the condition of a
finite mass spectrum was formulated as
\begin{equation}\label{Mfund1}
m \leq  M,
\end{equation}
where the maximal mass parameter $M$, called the fundamental mass,
was a new physical constant. The quantity $M$ was considered as
the curvature radius of a 5-dimensional hyperboloid whose surface
represented an implement-ation of the curved momentum 4-space, or
anti-de Sitter space,
\begin{equation}\label{O32}
    p_0^2 - p_1^2 - p_2^2 - p_3^2 + p_5 ^2 = M^2.
\end{equation}

It can be easily seen that for a free particle, $p_0^2 -
\overrightarrow{p}^2 = m^2$, condition (2) is automatically
satisfied on surface (3). It is also obvious that in the
approximation
\begin{equation}\label{Plpred}
    |p_0|,\;\;|\overrightarrow{p}| \ll M, p_5 \cong M
\end{equation}
anti-de Sitter geometry is transformed into Minkowski geometry in
the 4-dimensional pseudo-Euclidean $p$-space (the so called
"planar limit").

Thus, a new theory was constructed in anti-de Sitter space; in
this theory objects with masses larger than $M$ cannot be
considered as elementary particles, since no local fields
correspond to them [7--17].

It is important to note that the idea of a fundamental mass is
closely connected with the concept of a fundamental length,
\begin{equation}\label{L}
l= \hbar/Mc .
\end{equation}
Its physical meaning can be partly elucidated by comparing l and
the Compton wavelength of a particle, $\lambda_C = \hbar / mc$. It
can be seen from formula (2) that $\lambda_C$ cannot be smaller
than $l$. Since, according to Newton and Wigner [18], the
parameter $\lambda_C$ characterizes the size of the spatial region
in which a relativistic particle with the mass m can be localized,
it should be admitted that the fundamental length $l$ should
introduce into the theory a universal constraint on the precision
of spatial localization of elementary particles.

The idea of introducing a fundamental length as a new universal
constant with the dimension of length characterizing a typical
space--time scale was actively discussed in literature (see, e.g.,
[19--27]). The main stimulus for using this parameter was the hope
that $l$ would make it possible to get rid of ultraviolet
divergence. A simpler solution to this problem is known to be
found. Now the fundamental length appears again in the theory in
quite a different context of quantity, in a certain sense
complementary to the fundamental mass.

It should be noted that QFT models with a parameter similar to the
fundamental length $l$ turned out to be nonlocal. Returning to
Kadyshevsky's theory, we point out once again the consistent use
of the requirement of locality of this version of the quantum
theory. That is why the principle of local gauge symmetry can
still be used in describing an interaction. The key idea while
combining mass limit postulate (2) and the field locality
condition is that it is necessary to modify the very notion of the
field.

To illustrate the abovesaid, let us first consider the simplest
case of the real scalar field $\varphi(x)$. It is known that the
free Klein-Gordon equation for $\varphi(x)$ has the form
\begin{equation}\label{KG}
    (\square + m^2)\varphi(x) = 0,
\end{equation}
where $\square$ is the d'Alembert operator and $\hbar = c= 1$.

After standard Fourier transformations,
\begin{equation}\label{FT}
    \varphi(x) = \frac{1}{(2\pi)^{3/2}} \int e^{-i p_\mu x^\mu}\;\varphi(p)\;d\,^4
    p\;\;\;\;\; ( p_\mu x^\mu = p^0 x^0 - \emph{\textbf{p}} \emph{\textbf{x}}
    )
\end{equation}
we obtain the equation of motion in the 4-dimensional Minkowski
momentum space,
\begin{equation}\label{KGp}
    ( m^2 - p^2 ) \varphi(p) = 0, \;\;\;\;\;\; p^2 = p_0^2 -
    \emph{\textbf{p}}^2.
\end{equation}
From the geometric point of view, $m$ is the radius of the
4-hyperboloid,
\begin{equation}\label{ms}
    m^2 = p_0^2 -
    \emph{\textbf{p}}^2,
\end{equation}
on which the field $\varphi(p)$ is defined. Hyperboloids of type
(9) with an arbitrary radius can be placed in Minkowski space.
This means that formally, the modern QFT remains a perfect logical
scheme and its mathematical structure does not change up to
arbitrarily large quantum masses.

How can one modify the equation of motion in order to take into
account mass limit condition (2)? Following [2, 11], we replace
the 4-dimensional Minkowski momentum space used in the standard
QFT to the anti-de Sitter momentum space with constant curvature
implemented on the surface of the 5-hyperboloid,
\begin{equation}\label{ads}
    p^2_0 - \emph{\textbf{p}}^2 + p_5^2 = M^2.
\end{equation}

Let us assume that in $p$-representation the scalar field
$\varphi$ is defined on this surface, i.e., it is a function of
five variables $p_0, \emph{\textbf{p}}, p_5$  connected by
relation (\ref{ads}),
\begin{equation}\label{dads}
    \delta(p^2_0 - \emph{\textbf{p}}^2 + p_5^2 - M^2)\varphi(p_0, \emph{\textbf{p}},
    p_5).
\end{equation}
Here, the energy $p_0$ and the 3-dimensional momentum
$\emph{\textbf{p}}$ preserve their regular meaning, and relation
for the mass shell (9) is satisfied. In this case, condition (2)
holds for the considered field $\varphi(p_0, \emph{\textbf{p}},
p_5)$.

It is clear from Eq. (11) that the definition of one function
$\varphi(p_0, \emph{\textbf{p}}, p_5)$ of five variables $(p_\mu,
p_5)$  is equivalent to the definition of two independent
functions $\varphi_1(p)$ and  $\varphi_2(p)$ of the 4-momentum
$p_\mu$:
\begin{equation}\label{12}
    \varphi(p_0, \emph{\textbf{p}}, p_5) \equiv \varphi(p, p_5) = \left(%
\begin{array}{c}
  \varphi(p, |p_5|) \\
  \varphi(p, - |p_5|) \\
\end{array}%
\right) = \left(%
\begin{array}{c}
  \varphi_1(p) \\
  \varphi_2(p) \\
\end{array}%
\right), |p_5| = \sqrt{M^2 - p^2}.
\end{equation}

Note that the appearance of a new discrete degree of freedom
\be\label{p5}\epsilon  =  p_5/|p_5|=\pm 1\ee and the pair of field
variables is the characteristic feature of the developed theory.
Due to satisfaction of relation (9), Klein–Gordon equation (8)
should also be satisfied for the field $\varphi(p_0,
\emph{\textbf{p}}, p_5)$,

\begin{equation}\label{nf}
(m^2 - p_0^2 + \emph{\textbf{p}}^2 ) \varphi(p_0,
\emph{\textbf{p}}, p_5) = 0.
\end{equation}
This relation, however, does not reflect mass spectrum limit
condition (2). Also, it cannot be used for elucidation of the
field dependence on the new quantum number $\epsilon = p_5 /
|p_5|$, i.e., for determining the fields ${\varphi_1} (p)$ and
${\varphi_2} (p)$.  In order to take into account these
requirements and find the modified equation satisfying them, we
use relations (9) and (10) and obtain

$$
 m^2 - p_0^2 + \emph{\textbf{p}}^2
=  p_5^2 - M^2{\cos\mu}^2,
$$
where  $\cos \mu = \sqrt{1 - \frac{m^2}{M^2}}$. Thus, we write the
following instead of (14):
\begin{equation}\label{razlKG}
    (p_5 + M \cos \mu)(p_5 - M \cos \mu)\varphi(p, p_5) = 0.
\end{equation}
This equality holds if
\begin{equation}\label{NKG}
    (p_5 - M \cos \mu)\varphi(p, p_5) = 0 .
\end{equation}
It is natural to assume that (16) is the new equation of motion
for scalar particles. It follows from Eqs. (16) and (12) that
\begin{equation}\label{NKG12}
    \begin{array}{c}
      (|p_5| - M \cos \mu)\varphi_1(p) = 0, \\ \\
      (|p_5| + M \cos \mu)\varphi_2(p) = 0. \\
    \end{array}
\end{equation}
These equations satisfy the above requirements, and Eq. (14) holds
for $\varphi_{1,2}(p)$. Then, using (17), we obtain
\begin{equation}\label{122}
    \begin{array}{c}
      \varphi_1(p) = \delta(p^2 - m^2)\widetilde{\varphi}_1(p)\\ \\
      \varphi_2(p) = 0.\\
    \end{array}
\end{equation}
Thus, the free field $\varphi(p, p_5)$ defined in anti-de Sitter
momentum space (10) describes scalar particles with the mass $m$
satisfying the condition $m \leq M$. Note that the two-component
character of new field (12) is not manifested on the mass shell
(due to second equality (18)). It can play an important role in
the field interaction, i.e., beyond the mass shell.

Following [15], we use the Euclidean formulation of the theory
that appears in the case of analytic continuation to purely
imaginary energy values,
\begin{equation}\label{eup}
    p_0 \rightarrow ip_4.
\end{equation}
In this case, we consider de Sitter momentum space, rather than
anti-de Sitter space (10),
\begin{equation}\label{ds}
- p_n^2 + p_5^2 = M^2, \;\;\;\;\;\; n = 1,2, 3, 4.
\end{equation}
Obviously,
\begin{equation}\label{p5}
    p_5 = \pm\sqrt{M^2 + p_n^2}.
\end{equation}
If we use (20), the Euclidean Klein-Gordon operator $m^2 + p^2$
can be written, similar to (15), in the following factorized form:
\begin{equation}\label{factf}
m^2 + p_n^2 = (p_5 + M \cos \mu)(p_5 - M \cos \mu).
\end{equation}
It is clear that the nonnegative functional
\begin{equation}\label{action}
\begin{array}{c}
  S_0(M) = \pi M \times \\ \\
  \int \frac{d^4 p }{|p_5|}\left[\varphi^+_1(p)2M (|p_5| - M \cos
    \mu)\varphi_1(p) + \varphi^+_2(p)2M (|p_5| + M \cos
    \mu)\varphi_2(p)\right], \\
\end{array}
\end{equation}
\begin{equation}\label{123}
    \varphi_{1,2}(p) \equiv \varphi(p, \pm |p_5|),
\end{equation}
plays the role of the functional of action of the free Euclidean
field $\varphi(p, p_5)$. This action can be written in the form of
the 5-dimensional integral
\begin{equation}\label{5act}
    \begin{array}{c}
       S_0(M) = 2\pi M \times  \\ \\
       \int \varepsilon(p_5) \delta(p_L p^L - M^2)d^5 p \left[\varphi^+(p, p_5)2M (p_5 - M \cos
    \mu)\varphi(p, p_5) \right],
     \end{array}
\end{equation}
where
$$
   \emph{L} = 1, 2, 3, 4, 5
$$
and the following notation is introduced:
\begin{equation}\label{epsilon}
\varepsilon(p_5) = \frac{p_5 }{|p_5|}.
\end{equation}
The Fourier transform and the configuration representation play a
special role in this approach. First, note that in basic relation
(20), which defines de Sitter space, all components of the
momentum 5-vector are equivalent. Therefore, the expression
$\delta(p_L p^L - M^2)\varphi(p,  p_5)$,  which is now used
instead of (11), can undergo the Fourier transform,
\begin{equation}\label{FT2}
    \frac{2M}{(2\pi)^{3/2}}\int e^{-i p_K x^K}\delta(p_L p^L - M^2)\varphi(p,
 p_5) d^5\; p = \varphi(x, x_5), \;\; K, L = 1, 2, 3, 4, 5.
\end{equation}
Function (27), obviously, satisfies the following differential
equation in the 5-dimensional configuration space:

\begin{equation}\label{pent}
    \left(\frac{\partial^2}{\partial x_5^2 } - \square + M^2\right)\varphi (x, x_5) =
    0.
\end{equation}
Integration with respect to $p_5$ in (27) yields
\begin{equation}\label{FT3}
\begin{array}{c}
  \varphi(x, x_5) = \frac{2M}{(2\pi)^{3/2}}\int e^{i p_n x^n} \frac{d^4
p}{|p_5|}
  \left[e^{-i |p_5| x^5}\varphi_1(p) + e^{i |p_5| x^5}\varphi_2(p)\right],
  \\ \\
 \varphi^+(x, x_5) = \varphi(x, - x_5),  \\
\end{array}
\end{equation}
which implies
\begin{equation}\label{FT4}
\frac{i}{M}\frac{\partial\varphi(x, x_5)}{\partial x_5} =
\frac{1}{(2\pi)^{3/2}}\int e^{i p_n x^n} d^4 p
  \left[e^{-i |p_5| x^5}\varphi_1(p) - e^{i |p_5|
  x^5}\varphi_2(p)\right] .
\end{equation}
Four-dimensional integrals (29) and (30) transform the fields
$\varphi_1(p)$  and  $\varphi_2(p)$ into the configuration
representation. The inverse transformation has the following form:

\begin{equation}\label{IFT}
    \begin{array}{c}
      \varphi_1(p) = \frac{-i}{2M (2\pi)^{5/2}} \int e^{-i p_n x^n} d^4 x \left[\varphi(x, x_5)
      \frac{\partial e^{i |p_5|
      x^5}}{\partial x_5} -
      e^{i |p_5| x^5}\frac{\partial\varphi(x, x_5)}{\partial x_5}\right],\\ \\
      \varphi_2(p) = \frac{i}{2M (2\pi)^{5/2}} \int e^{-i p_n x^n} d^4 x \left[\varphi(x, x_5)
      \frac{\partial e^{- i |p_5|
      x^5}}{\partial x_5} -
      e^{ - i |p_5| x^5}\frac{\partial\varphi(x, x_5)}{\partial x_5}\right].  \\
    \end{array}
\end{equation}

Note that the independent field variables
\begin{equation}\label{phix}
    \varphi(x, 0)\equiv \varphi(x) = \frac{2M}{(2\pi)^{3/2}}\int e^{i p_n x^n} d^4\,p
  \frac{\varphi_1(p) + \varphi_2(p)}{|p_5|}
\end{equation}
and
\begin{equation}\label{chix}
    \frac{i}{M}\frac{\partial\varphi(x, 0)}{\partial x_5} \equiv \chi(x)=
\frac{1}{(2\pi)^{3/2}}\int e^{i p_n x^n} d^4 p
  \left[\varphi_1(p) - \varphi_2(p)\right]
\end{equation}
can be interpreted as the initial data for the Cauchy problem on
the surface $x_5 = 0$ for hyperbolic Eq. (28).

Now, substituting quantities (31) into action (23), we have
\begin{equation}\label{action2}
 \begin{array}{c}
  S_0(M) =  \frac{1}{2}\int d^4 \,x \left[\left|\frac{\partial \varphi(x, x_5)}{\partial x_n}\right|^2 +
    m^2|\varphi(x, x_5)|^2 + \left|i \frac{\partial \varphi(x, x_5)}{\partial x_5} - M \cos
    \mu \varphi(x, x_5)\right|^2\right] \equiv \\ \\
   \equiv \int L_0 (x, x_5)d^4 \,x .\\
 \end{array}
\end{equation}
It can be easily verified that, due to Eq. (28), action (34) is
independent of $x_5$,

\begin{equation}\label{ds0dx5}
    \frac{\partial S_0(M)}{\partial x_5} = 0.
\end{equation}
Therefore, the variable  $x_5$  can be arbitrarily chosen, and
$S_0(M)$ can be considered as a functional on the corresponding
initial data of the Cauchy problem for Eq. (28). For example, for
$x_5 =0$ we have

\begin{equation}\label{S0M}
\begin{array}{c}
    S_0(M) =  \frac{1}{2}\int d^4 \,x \left[\left(\frac{\partial \varphi(x)}{\partial x_n}\right)^2 +
    m^2(\varphi(x))^2 + M \left(\chi(x) - \cos
    \mu \varphi(x)\right)^2\right] \equiv \\ \\
   \equiv \int L_0 (x, M)d^4 \,x .
   \end{array}
\end{equation}
Thus, it was demonstrated that in this approach, the theory
preserves the property of locality; moreover, it becomes more
extended, covering the fifth coordinate $x_5$ as well.

The new density of the Lagrange function $L_0 (x, x_5)$ (see (34))
is a Hermitian form constructed from the fields $\varphi(x, x_5)$
and the components of the 5-dimensional gradient $\frac{\partial
\varphi(x)}{\partial x_L}, (L = 1, 2, 3, 4, 5).$ It is clear that,
although $L_0 (x, x_5)$ formally depends on $x_5$, the model essentially repeats
the 4-dimensional theory (see (35) and (36)).

It follows from the above transformations that the dependence of
action (36) on the two functional arguments $\varphi(x)$ and
$\chi(x)$  is directly connected with the fact that in the
momentum space the field has a doublet structure,
$\left(\begin{array}{c}
\varphi_1(p) \\
\varphi_2(p)
\end{array}\right) $, determined by
two signs of $p_5$. The Lagrangian $L_0 (x, M)$ however, does not
contain a kinetic term corresponding to the field $\chi(x)$.
Therefore, this variable is auxiliary.

A special role of the 5-dimensional configuration space in the new
formalism is also determined by the fact that its introduction
makes it possible to define the transformation of the local gage
symmetry of the theory. The object of these transformations is the
initial data in Eq.
\begin{equation}\label{id}
    \left(%
\begin{array}{c}
  \varphi(x, x_5) \\ \\
  \frac{i}{M} \frac{\partial\varphi(x, x_5)}{\partial x_5} \\
\end{array}%
\right)_{x_5 = fixed\;\; value} ,
\end{equation}
considered for fixed values of $x_5$.

Let us elucidate this point in more detail assuming that the field
$\varphi(x, x_5)$ is non-Hermitian and is associated with some
group of internal symmetry,

\begin{equation}\label{isg}
    \varphi' = U \varphi.
\end{equation}
Due to the local character of this group in the 5-dimensional
$x$-space,
\begin{equation}\label{lisg}
    U \rightarrow U(x, x_5),
\end{equation}
and the following gage transformations appear for initial data
(37) in the plane   $x_5 = 0$:
\begin{equation}\label{gt}
    \begin{array}{c}
      \varphi'(x) = U(x, 0)\varphi(x), \\ \\
      \chi'(x) = \frac{i}{M} \frac{\partial U(x, 0)}{\partial x_5}\varphi(x) + U(x, 0)\chi(x).\\
    \end{array}
\end{equation}
The group character of transformations (40) is quite obvious. The
explicit form of the matrix $U(x, x_5)$  can be determined from
the new theory of vector fields which is obtained from the
standard theory following the considered approach.

It is clear that Eq. (28) can be represented in the form of a
system of two first order equations with respect to the derivative
$\frac{\partial}{\partial x_5}$ [12],
\begin{equation}\label{fode}
    \left\{ \frac{i}{M} \frac{\partial }{\partial x_5} - \left[\sigma_3 \left(1 - \frac{\square}{2
    M^2}\right)-i\sigma_2 \frac{\square}{2
    M^2}
    \right]\right\} \phi(x, x_5) = 0,
\end{equation}
where
\begin{equation}\label{phixx5}
    \phi(x, x_5) = \left(%
\begin{array}{c}
  \frac{1}{2} \left[\varphi(x, x_5) +  \frac{i}{M}\frac{\partial\varphi(x, x_5) }{\partial
  x_5}\right]\\ \\
  \frac{1}{2} \left[\varphi(x, x_5) -  \frac{i}{M}\frac{\partial\varphi(x, x_5) }{\partial x_5}\right]\\
\end{array}%
\right) \equiv \left(%
\begin{array}{c}
  \phi_I(x, x_5) \\ \\
  \phi_{II}(x, x_5)\\
\end{array}%
\right),
\end{equation}
and $\sigma_i$, where  $i = 1, 2, 3$, are the Pauli matrices.
Comparing (42) with (32) and (33), one can find the relations
between the initial data of the Cauchy problem for Eq. (28) and
the solutions to system (41)
\begin{equation}\label{con}
    \phi(x, 0) = \left(%
\begin{array}{c}
  \phi_I(x, 0) \\ \\
  \phi_{II}(x, 0)\\
\end{array}%
\right) = \left(%
\begin{array}{c}
 \frac{1}{2}(\varphi(x) + \chi(x)) \\ \\
 \frac{1}{2}(\varphi(x) - \chi(x))  \\
\end{array}%
\right) \equiv \phi(x).
\end{equation}
It can be easily shown that in basis (43) the Lagrangian $L_0 (x,
M)$   (see (36)) has the following form:
\begin{equation}\label{L0MI}
L_0 (x, M) = \frac{\partial\phi(x) }{\partial
  x_n}(1 + \sigma_1)\frac{\partial\phi(x) }{\partial
  x_n} + 2 M^2\phi(x)(1 - \cos\mu\,\sigma_3)\phi(x).
\end{equation}

Let us consider the problem of transition from the new scheme to
the standard Euclidean QFT (the so called “correspondence
principle”). The 4-dimensional Euclidean momentum space, i.e., the
“planar limit” of the de Sitter momentum space, can be associated
with approximation (4),

\begin{equation}\label{FlatLp}
    \begin{array}{c}
      |p_n| \ll M \\
      p_5 \simeq M . \\
    \end{array}
\end{equation}
In the same limit in the configuration space we have
\begin{equation}\label{Flatlx}
\begin{array}{c}
  \varphi(x, x_5) = e^{-iM x_5}\varphi(x)  \\
  \chi(x) = \varphi(x) \\
\end{array}
\end{equation}
or
\begin{equation}\label{Flatlx1}
   \phi(x) = \left(%
\begin{array}{c}
  \varphi(x) \\
  0 \\
\end{array}%
\right) .
\end{equation}

Corrections of order $O(\frac{1}{M^2})$ to zero approximation (47)
can be easily obtained [13, 14] using (41),
\begin{equation}\label{1cor}
     \phi(x) = \left(%
\begin{array}{c}
  \left(1 - \frac{\square}{4 M^2}\right)\varphi(x) \\ \\
  \frac{\square}{4 M^2} \varphi(x) \\
\end{array}%
\right),
\end{equation}
which yields (see (43))
\begin{equation}\label{phi-chifl}
\varphi(x) - \chi(x) = \frac{\square \varphi(x)}{2 M^2}
\end{equation}
Taking into account (49) and (15), it can be concluded that in the
"planar limit" (formally, in the limit $M \rightarrow \infty$) the
Lagrangian $L_0(x, M)$ from (36) coincides with that of the
conventional Euclidean theory.

Since the new QFT is developed based on de Sitter momentum space
(20), it is natural to assume that in this approach the fermion
fields $\psi_\alpha(p, p_5)$ should be de Sitter spinors, i.e.,
they should be subject to the transformation with the
4-dimensional representation of SO(4, 1) group. Therefore,
hereinafter we use the basis of $\gamma$-matrices $(\gamma^4 =
i\gamma^0)$ in the form

\begin{equation}\label{gammam}
   \begin{array}{c}
     \gamma^L = (\gamma^1, \gamma^2, \gamma^3, \gamma^4, \gamma^5) \\ \\
     \left\{\gamma^L, \gamma^M\right\}  = 2 g^{LM},\\ \\
g^{LM} = \left(%
\begin{array}{ccccc}
  -1 & 0 & 0 & 0 & 0 \\
  0 & -1 & 0 & 0 & 0\\
  0 & 0 & -1 & 0 & 0 \\
  0 & 0 & 0 & -1 & 0 \\
  0 & 0 & 0 & 0 & 1 \\
\end{array}%
\right).
   \end{array}
\end{equation}
Obviously,
\begin{equation}\label{Dir}
\begin{array}{c}
   M^2 - p_L p^L = M^2 + p_n^2 - p_5^2 = ( M - p_L \gamma^L )( M + p_L \gamma^L
    ) = \\ \\
 = ( M + p^n \gamma^n - p^5 \gamma^5)(M - p^n \gamma^n + p^5
    \gamma^5).\\
\end{array}
\end{equation}
In the "planar limit" $M\rightarrow\infty$ the fields
$\psi_\alpha(p, p_5)$  become conventional Euclidean spinors.

It is clear that relations (27)--(33) considered for the scalar
field are also applicable in the fermion case. Let us give some
relations:

\begin{equation}\label{5FTF}
\psi(x, x_5) = \frac{2M}{(2\pi)^{3/2}}\int e^{-i p_K
x^K}\delta(p_L p^L - M^2)\psi(p,
 p_5) d^5\; p ,
\end{equation}
\begin{equation}\label{pentF}
    \left(\frac{\partial^2}{\partial x_5^2 } - \square + M^2\right)\psi(x, x_5) =
    0,
\end{equation}

\begin{equation}\label{psix}
\begin{array}{c}
  \psi(x, 0)\equiv \psi(x) = \frac{2M}{(2\pi)^{3/2}}\int e^{i p_n x^n} d^4\,p
  \frac{\psi_1(p) + \psi_2(p)}{|p_5|} = \\ \\
  = \frac{1}{(2\pi)^{3/2}}\int e^{i p_n x^n} \psi(p) d^4\,p\\
\end{array}
\end{equation}
\begin{equation}\label{chixF}
\begin{array}{c}
  \frac{i}{M}\frac{\partial\psi(x, 0)}{\partial x_5} \equiv \chi(x)=
\frac{1}{(2\pi)^{3/2}}\int e^{i p_n x^n} d^4 p
  \left[\psi_1(p) - \psi_2(p)\right] =  \\ \\
  = \frac{1}{(2\pi)^{3/2}}\int e^{i p_n x^n} \chi(p) d^4\,p
.\\
\end{array}
\end{equation}

Following Osterwalder and Schrader [28] \footnote {Note that in
[18] the so called Wick rotation is also interpreted in terms of
the 5-dimensional space.} we write the Euclidean fermion
Lagrangian in the form

\begin{equation}\label{EucFL}
\begin{array}{c}
  L_E(x)= \overline{\zeta}_E(x)\left(- i \gamma_n\frac{\partial}{\partial x^n} +
    m\right)\psi_E(x), \\ \\
  \left\{\gamma^n, \gamma^m\right\} = -2
    \delta^{nm}\;\;\; ( m,n = 1, 2, 3, 4).\\
\end{array}
\end{equation}
Here, the spinor fields $\overline{\zeta}_E(x) = \zeta^+_E
(x)\gamma ^4  $ and $\psi_E(x)$  are the independent Grassmann
variables which are not connected between each other by Hermitian
or complex conjugation. Correspondingly, the action is also
non-Hermitian.

It can be easily seen that the expression $2M(p_5 - M \cos \mu)$
which in our approach replaces the Euclidean Klein–Gordon operator
$p^2_n + m^2$ (see (36)), can be represented as

\begin{equation}\label{Dir3}
\begin{array}{c}
  2M(p_5 - M \cos \mu) =  \\ \\
  = \left[p_n\gamma^n - (p_5 -M)\gamma^5 + 2M sin \frac{\mu}{2}\right]
   \left[- p_n\gamma^n + (p_5 -M)\gamma^5 + 2M sin
   \frac{\mu}{2}\right] . \\
\end{array}
\end{equation}
In Euclidean approximation (45) relation (57) takes the form
\begin{equation}\label{fldir}
    p^2_n + m^2 = \left( p_n\gamma^n + m\right)\left(- p_n\gamma^n +
    m\right).
\end{equation}

Thus, the following expression can be used as a modified Dirac
operator:
\begin{equation}\label{Dirop}
    \mathcal{D}(p, p_5) \equiv p_n\gamma^n - (p_5 -M)\gamma^5 + 2M sin
    \frac{\mu}{2}.
\end{equation}

It is quite important that the new Klein–Gordon operator $2M(p_5-
M\cos\mu)$ can be expanded into matrix factors in another way
independent of (57),

\be \label{K4} 2M(p_5- M\cos\mu)= \big[\gamma^0 p_0+{\bf{\gamma
p}}
\end{equation}
$$
+\gamma^5(p_5+M)+  2 M {\cos}\mu/2\big]\big[\gamma^0
p_0-{\bf\gamma p}+\gamma^5(p_5+M)- 2M {\cos}\mu/2\big].
$$

Thus, in the approach under consideration we obtain some exotic
fermion field associated with the wave operator [1, 15] that has
no analogues in the common theory,

\be \label{K5} D_{exotic}(p,M)= p_\nu \gamma^\nu + (p_5+M)\gamma^5
-2M\cos(\mu/2). \ee The main difference of the operator
$D_{exotic} (p,m)$ from operator (59) is that it has no planar
limit (see (4)) and thus, it cannot serve for description of the
known particles. Therefore, (61) may correspond to the description
of\emph{ fermions unknown in SM.
}
The developed formalism [1] can be used to construct the
expression for the action of a fermion field in de Sitter momentum
space [15],
\begin{equation}\label{FErA}
    \begin{array}{c}
       S_0(M) = 2\pi M \int \varepsilon(p_5) \delta(p_L p^L - M^2)d^5 p\times  \\ \\
      \times \Big[\;\;\overline{\zeta}(p, p_5)\,\, \big[ p_n\gamma^n - (p_5 -M)\gamma^5
      + 2M\sin\frac{\mu}{2} \big] \,\,\psi(p, p_5)\;\Big]. \\
\end{array}
\end{equation}
If in the limit we change the variables
\begin{equation}\label{fvar}
    \begin{array}{c}
      \psi(p) = \frac{M}{|p_5|} (\psi(p, |p_5|) + \psi(p, -|p_5|)) \equiv M \frac{\psi_1(p) + \psi_2(p)}{|p_5|}\\ \\
      \chi (p) = \psi_1(p) - \psi_2(p)\\ \\
      \overline{\zeta}(p) = M \frac{\overline{\zeta}_1(p) + \overline{\zeta}_2(p)}{|p_5|} \\ \\
      \overline{\xi}(p) = \overline{\zeta}_1(p) - \overline{\zeta}_2(p),\\
    \end{array}
\end{equation}
representing the Fourier images of the local fields  $\psi(x),
\chi(x), \overline{\zeta}(x)$ and $\overline{\xi}(x) (x)  $
(compare with (54) and (55)), we obtain
\begin{equation}\label{S0D}
\begin{array}{c}
   S_0^{\mathcal{D}} = - \pi\int d^4 p \left(M + \frac{p^2_n}{M}\right)\overline{\zeta}(p)\gamma^5
   \psi(p)+\\ \\
  + \pi\int d^4 p\overline{\zeta}(p)\left(\slashed{p} + M\gamma^5 + 2M sin \frac{\mu}{2}\right)\chi (p)
  +\\ \\
  + \pi\int d^4 p \overline{\xi(p)}\left(\slashed{p} + M\gamma^5 + 2M sin \frac{\mu}{2}\right)\psi (p) -
  \\ \\
 - \pi\int d^4 p M \overline{\xi(p)}\gamma^5 \chi (p) . \\ \\
\end{array}
\end{equation}
In the configuration space we obtain
\begin{equation}\label{S0Dx}
    \begin{array}{c}
      S_0^{\mathcal{D}} = \int L_0^\mathcal{D}(x, M) d^4x =  \\ \\
      = \frac{1}{2} \int d^4 x \overline{\zeta}(x)\left(\frac{\square}{M^2} - 1\right)\gamma^5 \psi(x)
      +\\ \\
      + \frac{1}{2} \int d^4 x \overline{\zeta}(x)\left( i \gamma^n \frac{\partial}{\partial x^n} +
      M\gamma^5 +
      2M sin \frac{\mu}{2}\right)\chi (x) + \\ \\
       + \frac{1}{2} \int d^4 x \overline{\xi(p)}\left( i \gamma^n \frac{\partial}{\partial x^n} +
       M\gamma^5 +
      2M sin \frac{\mu}{2}\right)\psi (x) -  \\ \\
      - \frac{1}{2} \int d^4 x \overline{\xi(x)}\gamma^5 \chi (x).
    \end{array}
\end{equation}
Therefore, the modified Dirac Lagrangian $L_0^\mathcal{D}(x, M) $
represents a local function of the spinor field variables
$\psi(x), \chi(x), \overline{\zeta}(x)$ and $\overline{\xi} (x)$.
 It should be noted that here the analogy with the boson case
(see (36)) is obvious.

Introducing the notation \footnote{Note that a similar notation
for the masses was used in [29].} \be\label{NEZN} m_1=2M\sin\mu/2,
m_2=2M\sin^2\mu/2,   m_3 = 2M\cos\mu/2,  m_4 = 2M\cos^2\mu/2, \ee
and $ p_\mu =i\partial_\mu $,   and going over to the Hamiltonian
form of the Dirac equations of motion, we can write

\be \label{DM}\Big(p_0 - {\bf\hat \alpha p}- \hat\beta m_1 -
\hat\beta\gamma^5 m_2\Big )\Psi(x,t,x_5)=0,\ee

\be \label{DM1}\Big(p_0 - {\bf \hat\alpha p}- \hat\beta { m_3}
-\hat \beta\gamma^5 { m_4}\Big )\Psi^{exotic}(x,t,x_5)=0. \ee

In these modified Dirac equations, the matrices $\hat\beta
=\gamma_0$, $\gamma^i=\hat\beta\hat\alpha^i$. \footnote{ It is
important to note that on the mass shell $p_5=M cos \mu$ there do not exist any
operators acting on the coordinate $x_5$, and this parameter can
be taken equal to zero without losing generality [15, 17].}

In the quantum mechanical approximation, the Hamiltonians
corresponding to Eqs. (67) and (68) can be represented in the
following form:

\be\label{H1} {\hat{H}} =
\overrightarrow{\hat{\alpha}}\overrightarrow{p} + \hat{\beta
}\left(m_1 + m_2\gamma_5\right), \ee

\be\label{H2} \hat{H}_{exotic} =
\overrightarrow{\hat{\alpha}}\overrightarrow{p} + \hat{\beta
}\left(m_3 + m_4\gamma_5\right). \ee

Apparently, expressions (69) and (70) turn out to be non-Hermitian
due to the $\gamma_5$-mass terms ( $H\neq H^{+}, H_{exotic}\neq
H^{+}_{exotic}$). Thus, the following conclusion can be made: mass
spectrum constraint (2) that is the basis of the geometric
approach to development of the modified QFT with a maximal mass
[15, 17] results in the appearance of non-Hermitian contributions
to Hamiltonians (69), (70).

\section{The theory with a limited mass as an algebraic
non-Hermitian $\cal PT$-symmetric theory}

The non-Hermitian quantum theory studying models with
non-Hermitian Hamiltonians has known great development in the
recent years [30-68]. The central problem of such theories is the
construction of a new scalar product in which the average values
of non-Hermitian Hamiltonians become real. This theory has become
so popular in a short time that it is now impossible to cite all
publications concerning this topic. They contain papers devoted to
the investigation of purely mathematical issues of this theory
(see, e.g., [30-34]). Some papers are devoted to the examination
of model Lagrangians and illustration illustration of the
capabilities of this method (for example, [35–38]). Studies in the
field of experimental physics are also present. Among those, the
most promising are the papers devoted to the application of the
pseudo-Hermitian approach in the field of nonlinear optics
[39.46]. Of interest is an attempt to apply this theory to
investigation of the problem of non-Hermitian interpretation of
the fundamental length [47, 48]. Since [47, 48] consider
non-relativistic Hamiltonians, this attempt is somewhat naive from
the point of view of relativistic quantum physics. Nonetheless, it
is important to underline that non-Hermitian Hamiltonians can be
considered as a certain fruitful medium for the search of new
physics beyond the Standard Model.

One of the variants in constructing a new scalar product is
implemented in  $\cal PT$-symmetric theories, i.e., models in
which the Hamiltonian possesses the combined  $\cal PT$-symmetry,
rather than separate  $\cal P$ and  $\cal T$ symmetries. This was
achieved by finding a special operator $\cal C$ which in some
sense can be associated with the charge conjugation operator, via
recurrence relations (see [57, 58]), and constructing with the
help of this operator a new scalar product.

Another method for constructing the operator is implemented in
pseudo-Hermitian theories [32]. These theories consider the models
with non-Hermitian Hamiltonians which have a pseudo-Hermitian
character,
\begin{equation}\label{eta0}
\eta_0 H \eta_0 ^{-1} = H^\dagger\label{ps},
\end{equation}
where $\eta_0 $ is a linear Hermitian operator. The operator
${\cal C}$ is constructed using $\eta_0 $ in the following way: $
{\cal C}= {\eta_0}^{-1} {\cal P}$ where ${\cal P}$ is the operator
of spatial reflection.

Among many models considered in the context of development of the
non-Hermitian quantum theory [30–68], there exists the so called
${\cal PT}$-${\cal PT}$-symmetric massive Thirring model [58],
developed in the space $(1+1)$ with the Hamiltonian density
\begin{equation}\label{B1} {\cal
H}(x,t)=\bar{\psi}(x,t)\Big(-i\overrightarrow{\partial}\overrightarrow{\gamma}+m_1+\gamma_5
m_2  \Big)\psi(x,t). \end{equation}

Generalizing the expression for the Hamiltonian density to the (3
+ 1)-dimensional case and writing the Hamiltonian following from
(72) in the form
\begin{equation}\label{B3}
H(x,t)=\gamma_0 \overrightarrow{\gamma}\overrightarrow{p}
+\gamma_0 (m_1+\gamma_5 m_2),
\end{equation}
we obtain for the equations of motion \be\label{B4}
\Big(i\partial_\mu\gamma^{\mu}-m_1-\gamma_5 m_2 \Big) \psi(x,t)=0.
\ee

It can be easily seen that the so-called physical mass $m$
appearing in this model as \begin{equation}\label{m} m^2= m_1^2 -
m_2^2,\end{equation} is real if the following inequality is
satisfied:
 \be \label{e210}
m_1^2\ge m_2^2. \ee This inequality was considered in this theory
as the basic requirement defining the region of unbroken $\cal
PT$-symmetry of the studied Hamiltonian. In [58] the operator
$\cal C$ providing for the modified scalar product was calculated
recurrently.

Obviously (see [69–76]), the model considered in [58] is similar
to Kadyshevsky’s model in its fermion sector [1, 2, 11, 15, 16]
from the point of view of the algebraic approach to non-Hermitian
models. Indeed, Hamiltonians (69), (70), and (73), as well as
equations of motion (67), (68), and (74) following from these
Hamiltonians, coincide to notation. In other words, a certain
analogy of Kadyshevsky’s model from the point of view of the
algebraic approach to non-Hermitian $\cal PT$-symmetric theories
was considered in [58]. The authors of [58] probably did know at
that time that the non-Hermitian $\gamma_5$-mass extension had
already been used by Kadyshevsky earlier. Moreover, it can be
easily seen that the approach developed in [58] was not logically
complete from the point of view of application to physics.

 In particular, the
following question arises while analyzing the model with
Hamiltonian (73): how can a particular physical particle be
described using this Hamiltonian? In other words, how can the
parameters $m_1$, $m_2$, for the particle description in the
framework of the non-Hermitian algebraic model can be found from
the known physical mass $m$ of this particle? Obviously, this
question cannot be answered unambiguously using conditions (75),
(76) alone. Equation (75) yields an infinite set of pairs $m_1$
and $m_2$ for the given mass $m$. The matter is that the model
defined by Hamiltonian (73) (or, (69), which is the same), is
two-parametric. The use of the only parameter $m$ is an attempt to
pass from the two-parametric approach to the one-parametric
description of this model. It is clear that such a “change of
variables” is ambiguous. In order to describe a physical system
using the parameter $m$ it is necessary to introduce the second
parameter. It is quite obvious that the choice of this parameter
should be dictated by physical considerations. Therefore, the
algebraic theory should also contain the parameter $m_{max}$
corresponding to the parameter $M$ in the geometric model.

It can be assumed that, similar to the parameter $M$ in the
geometric theory, the parameter $m_{max}$ should be a mass
limiting parameter for the algebraic model. Some suggestive
considerations can be easily obtained in support of this
statement. Indeed, using the theorem about the arithmetic mean and
geometric mean of two numbers and  of two numbers  $m^2$ and
$m_2^2$, we have (see, e.g., [73])
 \be \label{AG}
   \frac{m^2
+{m_2}^2}{2} \geq \sqrt{m^2 {m_2}^2}, \ee which, taking into
account (75), yields the following inequality
 \be \label{M} m\leq
\frac{{m_1}^2}{2 m_2} \equiv m_{max}. \ee Note that here
constraint (78) is formal yet, since the value of $m_{max}$ is
determined by the parameters $m_1$  and $m_2$ of the theory whose
values in the general case can vary infinitely. It will be shown
below that a closer connection between $m_{max}$ and the
fundamental mass $M$ of the geometric theory can be established.
For this purpose it is sufficient, similar to what was done in
Kadyshevsky’s geometric model, to postulate the existence of the
maximal mass parameter equal to $M$. The necessity of introducing
this postulate will become clear below.

The correctness of the developed algebraic approach is verified by
the following: in the Hermitian limit $m_2 \rightarrow 0$ we
obtain from (78) \be \label{inf} m_{max} \rightarrow \infty ,\ee
which corresponds to the transition to the standard Dirac theory
in which there is no constraint on the fermion mass. Therefore,
limit (79) is not only correct but also means that the considered
algebraic model satisfies the correspondence principle, i.e., in
this limit it is transformed into the standard Dirac model. In
this sense, in “planar limit” (4), in which formally $M
\rightarrow \infty$ we obtain the transition from the curved de
Sitter momentum space to Minkowski space and also obtain the
Hermitian limit [69].

It can be easily seen that conditions (75), (76), and (78) are
automatically satisfied if we introduce the following
parameterization [69] following from the solution to the system of
equations

\begin{equation}\label{arr1}
 \left\{
   \begin{array}{rcl}
      m_1^2  -  m_2^2  = & m^2; \\
     \frac {m_1^2}{2 m_2}  = & \,\,\, m_{max}, \\
    \end{array}
           \right.
\end{equation}

namely,

\be\label{m11} m_1^{\mp} = \sqrt{2} m_{max}\sqrt{1 \mp
\sqrt{1-m^2/m^2_{max}}};
 \ee

\be \label{m22}m_2^{\mp} = m_{max}\left( 1 \mp
\sqrt{1-m^2/m^2_{max}}\right). \ee It can be seen from Eqs. (81)
and (82) that $m_1$ and $m_2$  are double-valued functions of the
physical mass $m$. To illustrate this double-valued character, we
give the plot characterizing the connection of the functions
$m_1^{\mp}$, $m_2^{\mp}$ and $m$  (see Fig. 1). Let us define the
reduced masses as follows: $\nu = m/m_{max}$, ${\nu}_1 =
m_1/m_{max}$ and ${\nu}_2 = m_2/m_{max}$.  Then we obtain from
(81), (82) \be\label{m1} \nu_1^{\mp} = \sqrt{2} \sqrt{1
\mp\sqrt{1-\nu^2}};
 \ee

\be \label{m2}\nu_2^{\mp} = 1 \mp\sqrt{1-\nu^2}. \ee

Figure 1 shows the parameters $\nu_1^{\mp}$, $\nu_2^{\mp}$  as
functions of $\nu$ [69], [70]. The region of existence of $\cal
PT$ - symmetry is now obvious, $0 \leq\nu\leq 1$. For these values
of $\nu_1$ and $\nu_2$  the modified Dirac equation with a maximal
mass describes the propagation of particles with real masses.

\begin{figure}[h] \vspace{-0.2cm} \centering
\includegraphics[angle=0, scale=0.5]{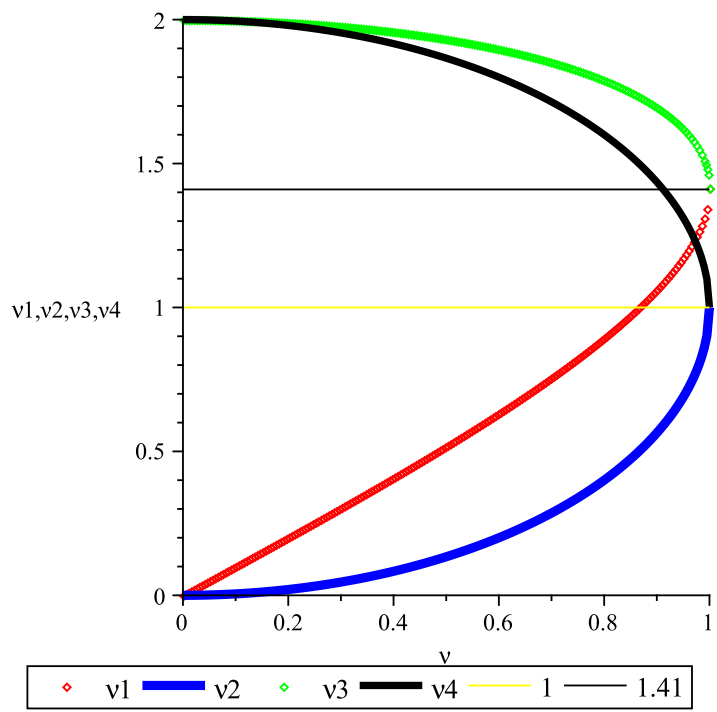}
\caption{Parameters $\nu_1^{\mp}, \nu_2^{\mp}$ as functions of
$\nu$.} \vspace{-0.1cm}\label{f4}
\end{figure}

In order to understand the physical meaning of the double-valued
dependence of $m_1$ and $m_2$  on $m$, $m_{max}$ we consider the
geometric theory with a limited mass. Substituting into formula
(66) defining the masses $m_1, m_2, m_3, m_4$, the value $\cos
    \mu = \sqrt{1 - \frac{m^2}{M^2}}$, we
obtain

\be\label{M1} m_1 = \sqrt{2} M\sqrt{1 - \sqrt{1-m^2/M^2}} ; \ee

\be\label{M2} m_2 = M \left( 1 - \sqrt{1-m^2/M^2}\right); \ee

\be\label{M3} m_3 = \sqrt{2} M\sqrt{1 + \sqrt{1-m^2/M^2}};  \ee

\be\label{M4} m_4 = M \left( 1 + \sqrt{1-m^2/M^2}\right). \ee

It is quite obvious that expressions (85)--(88) and (81), (82)
coincide to the change of variable  $M \leftrightarrow m_{max}$:

$$\frac
{m_1^2}{2m_2}= \frac {m_3^2}{2m_4}=  M \longleftrightarrow
 \frac {{m_1^\mp }^2}{2m_2^\mp }= m_{max}.$$

This means that if the parameter $m_{max}$ is introduced in the
algebraic theory and identified with the maximal mass $M$ from
Kadyshevsky’s geometric theory, it turns out that the algebraic
model also contains the description of “exotic” particles, which
was earlier considered to be the capability of the geometric
approach. In other words, it can be established that, similar to
the parameters $m_1$  and $m_2$  in the geometric theory which
participate in description of ordinary particles, the parameters
$m_1^-$, $m_2^-$ play the same role in the algebraic model (they
correspond to the lower branches of the plots, in Fig. 1).
Similarly, the parameters $m_3$, $m_4$, are used for description
of exotic particles in the geometric model, and in the algebraic
model the corresponding parameters are $m_1^+$, $m_2^+$  (upper
branches of the plots, $\nu_1^+$,  $\nu_2^+$). The regions of
variation of the physical mass $m$ and the parameters $m_1$ and
$m_2$ are

\be\label{llee}
0 \leq m \leq m_{max}; \qquad m \leq m_1 \leq 2 m_{max}; \qquad 0
\leq m_2 \leq 2 m_{max}. \ee They correspond to the following
regions on the plot, respectively: $$0 \leq \nu \leq 1;\qquad \nu
\leq \nu_1 \leq 2; \qquad 0 \leq \nu_2 \leq 2. $$

 Apparently, in the algebraic interpretation these
constraints define the region of unbroken $\cal PT$ -symmetry of
the model corresponding to (76). The point  $m=m_{max}$  ($\nu=1$
on the plot) is a special case: it corresponds to the maximon. At
this point of the plot we have $\nu_1^-=\nu_1^+=\sqrt{2}$ and
$\nu_2^-=\nu_2^+=1$. Note that with the appearance of exotic
particles the theoretical essence of the maximon does not change
either physically or mathematically. It still plays the role of
the particle with the maximal mass. Note that in the geometric
model the appearance of “exotic” particles was considered to be a
consequence of the approach itself in which the new unusual
particle properties were connected with the new degree of freedom
in the theory, the sign of the momentum component $p_5$ ($\epsilon
= p_5/|p_5|=\pm 1$ (see [2, 11]). It can be seen, however, that in
the algebraic approach the introduction of the parameter $m_{max}$
also makes it possible to include the description of “exotic”
particles in the theory. Thus, the appearance of “exotic”
particles is directly connected with the non-Hermitian character
of the considered Hamiltonian [69], [70].

This also makes it possible to specify the region of $\cal
PT$-symmetry of the model. It can be seen from Fig. 2 that the
region of $\cal PT$-symmetry of the Hamiltonian \be\label{on}
{\hat{H}} = \overrightarrow{\hat{\alpha}}\overrightarrow{p} +
\hat{\beta }\left(m_1^{\mp} + m_2^{\mp}\gamma_5\right) \ee  in the
plane $m_1, m_2$ is determined by three groups of inequalities
(with account of the possible change of sign of the parameter
$m_2$):
$$I.\,\,\,\,\,\,\,\,\,\,\,\,\,\,\,\,\,\, m_1/\sqrt{2} \leq m_2 \leq m_1,$$
$$II. -m_1/\sqrt{2}\leq m_2\leq m_1/\sqrt{2},$$
$$III. \,\,\,\,\,\,\,\, -m_1 \leq m_2\leq -m_1/\sqrt{2}.$$
It should be underlined that, unlike [58], here the region of
$\cal PT$-symmetry is defined in detail and it is demonstrated
that while the central subregion $II$ corresponds to an ordinary
particle, subregions $I$ and $III$ correspond to exotic particles.
Thus, it is absolutely clear that expression (76) cannot be
considered as the only constraint in the theory, and introducing
the parameter $m_{max}$ and taking into account inequality (78)
makes it possible to specify more precisely the region of $\cal
PT$-symmetry of the model [70].

\begin{figure}[h]
\vspace{-0.2cm} \centering
\includegraphics[angle=0, scale=0.5]{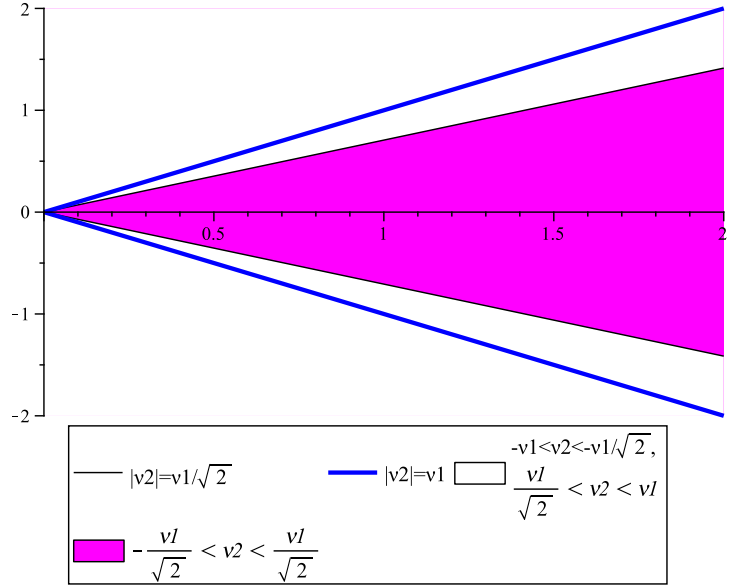}
\caption{The parametric region of unbroken $\cal PT$-symmetry
${m_1}^2 \geq {m_2}^2$  for Hamiltonian (90)
consists of three subregions. Shaded subregion $II$ corresponds to
ordinary particles, and two neighboring subregions $I$ and $III$
correspond to exotic fermions.} \vspace{-0.1cm}\label{f4}
\end{figure}

The following fact is surprising: in the algebraic model for any
pre-set values of $m_1$ and $m_2$, with account of (80),
constraint (78) appears automatically, similar to the geometric
theory. Indeed, let us consider the parameter [72] \be
\label{xim1} \xi = \frac {m_1}{M}= \frac {2m_2}{m_1}. \ee Taking
into account (80), we have \be \label{xim2} \frac {m_2}{M}= \frac
{\xi^2}{2}. \ee Figure 3 shows the dependence of the quantities
$m/m_1$, $M/m_1$ and  $m/M$  on the parameter $\xi =  \frac
{m_1}{M}= \frac {2m_2}{m_1}.$ In particular, it can be seen that
the function $m/M$ has a maximum at the point $m=M$. The maximal
particle mass $m=M$  is achieved when the auxiliary mass
parameters satisfy the following relation: $m_2 = m_1/\sqrt 2.$
One can find the parameters $m_1$ and $m_2$  for which there
exists the limiting transition to the standard Dirac equation up
to this value of the parameter. Higher values of $m_2$ bring us to
the decreasing branch of the curve $m/M$, and there is no Dirac
limit in this region, while at the point $m_1 = m_2=2m_{max}$  the
value of $m$ is again equal to zero. Similar to Fig. 1, this means
that now the case of massless particles, for example, corresponds
to the two points: $m_1=m_2=0 $  and $ m_1=m_2=2 M$.  In the first
case, we have the description of ordinary massless fermions, and
the second case should be interpreted as the description of their
exotic partners. This yields that the region $m_1 < \sqrt 2 m_2$
corresponds to the description of “exotic” particles for which
there is no transition to the Hermitian limit. In this case, the
point $m_2 = m_1 = 2M$  corresponds to massless exotic fermions.

\begin{figure}[h]
\vspace{-0.2cm} \centering
\includegraphics[angle=0, scale=0.5]{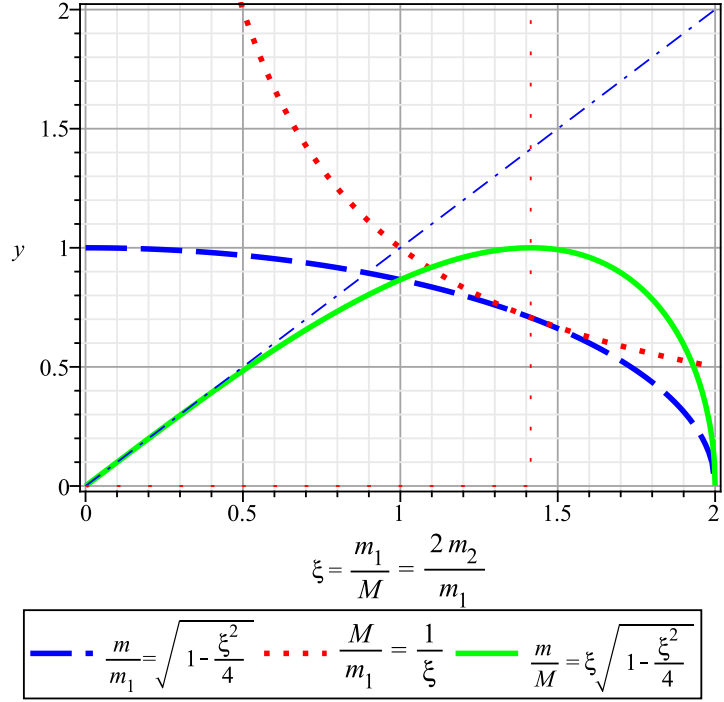}
\caption{Ratios $m/M$, $M/m_1$ and $m/m_1$ as functions of $\xi =
\frac {m_1}{M}= \frac {2m_2}{m_1}.$} \vspace{-0.1cm}\label{f5}
\end{figure}

Expressions (81), (82) make it possible to consider the “physical”
approach in this algebraic model [75], [76], i.e., to answer the
question posed earlier: how can a particular particle be described
using Hamiltonian (69)? In other words, how can one find the
parameters $m_1$, $m_2$ for the particle description in the
framework of the algebraic model using the known physical mass $m$
of the particle? It was already noted that it is impossible to
answer this question unambiguously using condition (75) alone. If
the parameter  $m_{max}$ is introduced, the answer follows from
Eqs. (81), (82). Actually, by doing so, the transition is made
from the two-parametric problem defined by Hamiltonian (69) with
the parameters $m_1$, $m_2$, to the two-parametric problem with
the parameters $m$, $m_{max}$. This proves once more the necessity
of introducing the parameter $m_{max}$, and taking into account
(78).

Let us consider the algebraic model describing the whole fermion
spectrum [70–76] in this context. The question on the unique
character of for all particles arises. It can be assumed from
physical considerations that the model in which $m_{max}$ is
unique for all particles is more preferable. Then it is logical to
conclude that this unique $m_{max}$, is the maximon mass and
should correspond to the limiting mass $M$ in the geometric
approach. This turns out to be possible because, as it was already
indicated above, the Hamiltonians and evolution equations in these
two models coincide to notation. In other words, it is necessary
to assume the equality of $m_{max}$ and $M$ in the theory
describing the mass spectrum of elementary particles from physical
considerations. Since the value of $M$ should follow from
experiment, the postulate of the equality of $m_{max}$ and $M$
results in the physical constraint on the mass spectrum
$m_{max}=M$, along with the condition $m_{max}=\frac
{m_1^2}{2m_2}$ in the algebraic model.

Thus, the connection of this algebraic model and the geometric
theory with a maximal mass is established once again [1, 2, 7–17].

\section{Scalar product in pseudo-Hermitian model}

Since the considered Hamiltonian is non-Hermitian
(pseudo-Hermitian, see (71)), it is necessary to introduce a new
scalar product defined by the operator  $\cal C$. For this purpose
we rewrite the mass term of the Hamiltonian in the form \be
\hat\beta(m_1 + m_2\gamma_5)=\hat{\beta}m( \mathrm{ch}
\alpha+\gamma_5 \mathrm{sh}
\alpha)=\hat{\beta}m\exp{(\gamma_5\alpha)}, \ee where $\mathrm{ch}
\alpha = m_1/m.$ Now we can write the original Hamiltonian as
follows: \be \label{H} \hat{H}= \hat{\overrightarrow{\alpha}}
\overrightarrow{p}+\hat{\beta}m\exp({\gamma_5\alpha}), \ee and the
Hermitian conjugate transpose Hamiltonian as \be
\hat{{H^\dagger}}= \hat{\overrightarrow{\alpha}}
\overrightarrow{p}+\hat{\beta}m\exp({-\gamma_5\alpha}). \ee Using
the commutation rules for the matrices ${\gamma}_5$,
$\hat{\vec{\alpha}}$ and $\hat{\beta}$ it can be easily shown that

\be\label{H}
    e^{\alpha\gamma_5 /2} {\hat{H}} = {\hat{H_0}}e^{\alpha\gamma_5 /2} = {\hat{H_0}}
    \eta
\ee and \be\label{H+}
    e^{-\alpha\gamma_5 /2} \hat{H^\dagger} ={\hat{H_0}}e^{-\alpha\gamma_5 /2} = {\hat{H_0}}
    {\eta}^{-1},
\ee where the following notation:
$$\hat H_0 =
\overrightarrow{\hat{\alpha}}\overrightarrow{p} + \hat{\beta }m,
$$
corresponding to the original free Dirac Hamiltonian is
introduced, and \be \label{eta} \eta = e^{\alpha\gamma_5 /2}.\ee
It can be easily seen from (94) and (97) that the Hermitian
Hamiltonian $\hat H_0$  and the Hamiltonians $\hat H$, $\hat
H^\dagger$ are connected by the non-unitary transformation $\eta$.
(See the geometric theory in de Sitter space [77] where a similar
transformation is unitary, for comparison).

It is also obvious that \be \eta_0 {\hat H} {\eta_0}^{-1} = {\hat
H^\dagger}, \ee  where the operator $\eta_0 = e^{\alpha \gamma_5}=
{\eta}^2$  defines the pseudo-Hermitian properties of the
Hamiltonian. Following [32], we can define the following operator:
\be \label{eta0} {\cal C}= {\eta_0}^{-1} {\cal P} = e^{-\alpha
\gamma_5} \gamma_0.\ee Using the standard representation of
$\gamma$-matrices, the operator ${\cal C}$ can be written in the
matrix form, \be \label{C1}{\cal C}=\left(\begin{array}{cc}0& I
\frac{m_1^{\mp}-m_2^{\mp}}{m}\\I \frac{m_1^{\mp} +
m_2^{\mp}}{m}&0\end{array}\right), \ee  where $
I=\left(\begin{array}{cc} 1 & 0 \\ 0 & 1 \end{array}\right)$. Note
that the operator ${\cal C}$ ${\cal C}$  in notation (71) or (101)
has a simpler form and is more convenient than the corresponding
expression in [58], since here (see [71, 76]) it is obtained
explicitly, while in [58] this operator was constructed via the
iteration method and written in the integral form.

It can be easily proved that the scalar product constructed using
${\cal C}$-operator (101) in a standard way [32] is positive
definite. Indeed (see [71, 76]), let us write an arbitrary vector
in the modified theory,
$$
\widetilde{\psi}= \left( \begin{array}{cc}
 x+i y&{} \\
u+iv&{}\\
z+iw&{}\\
t+ip&{}
\end{array}\right).
$$
It can further be proved by direct calculations using
representation (101) for the operator ${\cal C}$, that  the norm
of the vector $\widetilde{\psi}$ in the considered theory with the
modified scalar product is given by the expression
 \be \label{Psi} \langle{}
\overline{\widetilde{\psi}} {\cal C}
|\widetilde{\psi}\rangle=\frac{m_1+m_2}{m}(x^2+y^2)+\frac{m_1+m_2}{m}(u^2+v^2)+
\frac{m_1-m_2}{m}(z^2+w^2)+\frac{m_1-m_2}{m}(t^2+p^2). \ee This
expression is positive definite in region (76) of ${\cal
PT}$-symmetry of the considered theory, i.e., for $m_1\geq m_2$.

It can also be easily seen that the average values of Hamiltonian
(90) in the modified scalar product are real. Let us write
 \be \label{mid}
\langle{} \overline{\widetilde{\psi}} {\cal C}\hat{H}
|\widetilde{\psi}\rangle=\langle{} \overline{\widetilde{\psi}}
e^{-\alpha \gamma_5}\gamma_0  \bigl(
\overrightarrow{\hat{\alpha}}\overrightarrow{p} + \hat{\beta }m
e^{\alpha \gamma_5} \bigr) \widetilde{\psi}\rangle .\ee Taking the
Hermitian conjugate transpose and performing elementary
commutations similar to (94), (97), we prove that this expression
is real.

Let us consider the problem of finding eigenvalues and
eigenvectors of Hamiltonian (69) [71],
$$\hat{H}\widetilde{\psi}=E\widetilde{\psi}.$$
Using the standard representation of $\gamma$-matrices, we write
$\hat{H}$ in the matrix representation,
$$\hat{H}=\left(%
\begin{array}{cccc}
  m_1 & 0 & p_3-m_2 & p_1-ip_2 \\
  0 & m_1 & p_1+ip_2 & -m_2-p_3 \\
  m_2+p_3 & p_1-ip_2 & -m_1 & 0 \\
  p_1+ip_2 & m_2-p_3 & 0 & -m_1 \\
\end{array}%
\right),$$ where $p_i$ are the momentum components. The condition
$$\det{(\hat{H}-E)}=(-E^2 + {m_1}^2-{m_2}^2 + {p_{\bot}}^2
+{p_3}^2)^2 =0$$ makes it possible to determine the eigenvalues of
the energy $E$,
 \be\label{E}
\emph{E}=\pm\sqrt{{m_1}^2-{m_2}^2 + {p_{\bot}}^2 +{p_3}^2},\ee
where $p_{\bot}=\sqrt{{p_1}^2+{p_2}^2}$ and ${m_1}^2-{m_2}^2 =m^2
$. It can be seen that the values of $E$ coincide with the
eigenvalues of the Hermitian operator $\hat{H_0}$.

Let us consider the wave function describing a free particle with
a certain energy and momentum. It can be represented as a plane
wave,
 \be\label{psi} \widetilde{\psi}=
\frac{1}{\sqrt{2E}}\widetilde{u}e^{-ipx}. \ee Here, the amplitude
$\widetilde{u} $ is a bispinor and satisfies the algebraic
equations \be\label{u} \left(\gamma{p}-m
e^{\gamma_5\vartheta}\right)\widetilde{u}=0; \ee

\be\label{u1} \overline{\widetilde{u}} \left(\gamma{p}-m
e^{-\vartheta\gamma_5}\right)=0, \ee where
$\overline{\widetilde{u}}={\widetilde{u}}^{*}\gamma_0$. The
solutions to Eqs. (106), (107) can be represented in the form (see
[71, 76])
\be\label{u2} \widetilde{u}=\sqrt{2m}\left(%
\begin{array}{c}
  A_1 w \\
  A_2 w \\
\end{array}%
\right); \ee

\be \label{u2}\overline{\widetilde{u}}=\sqrt{2m}\left(%
\begin{array}{cc}
 A_1^{*} w^{*}, & - A_2^{*} w^{*} \\
\end{array}%
\right),\ee where $A_1$,  $A_2$ are defined by the expressions
$$
A_1=\mathrm{ch}\frac{\alpha}{2}\mathrm{ch}\frac{\beta}{2}
+\mathrm{sh}\frac{\alpha}{2}\mathrm{sh}\frac{\beta}{2}(\textbf{n}
\overrightarrow{{\sigma}}) ;
$$
$$
A_2= \mathrm{sh}\frac{\alpha}{2}\mathrm{ch}\frac{\beta}{2}
+\mathrm{ch}\frac{\alpha}{2}\mathrm{sh}\frac{\beta}{2}(\textbf{n}
\overrightarrow{{\sigma}}).
$$

Here, $\mathrm{ch} \,\alpha = m_1/m$,\,\, $\mathrm{sh} \,\alpha =
m_2/m$  and $\mathrm{ch} \,\beta=\emph{\emph{\emph{E}}}/m$,\,\,
$\mathrm{sh} \,\beta=p/m$, and $w $ is the two-component spinor
satisfying the condition
\be\label{w}
           \overrightarrow{\sigma}{\textbf n}w_{\zeta}=\zeta w_{\zeta}
\ee
and the normalization relations
$$
   w^{*} w =1.
$$
We recall here the standard notation: $\overrightarrow{\sigma}$
are the Pauli matrices with the dimension $2\times 2$, and
$\textbf{n}=\textbf{p}/p $ is the unit vector along the momentum.

Solving Eq. (110), we have
$$
   w_{1} = \left(%
\begin{array}{c}
  e^{-i\varphi/2}\cos\theta/2 \\
  e^{i\varphi/2}\sin\theta/2 \\
\end{array}%
\right),\,\,\,\,\,                       w_{-1} = \left(%
\begin{array}{c}
  -e^{-i\varphi/2}\sin\theta/2 \\
  e^{i\varphi/2}\cos\theta/2 \\
\end{array}%
\right),
$$
where $\theta$ and $\varphi$  are the polar and azimuthal angles
defining the direction of \textbf{n} relative to the axes
$x_1,x_2,x_3$.

Satisfaction of the following condition can be easily verified by
direct calculation:
$$
\overline{\widetilde{u}}\widetilde{u} = 2 m.
$$
This result is quite expected, since there exists the
transformation connecting the bispinor amplitudes of the modified
equations $\overline{\widetilde{u}},\widetilde{u}$ and the
corresponding solutions to the original Dirac equations,
$$ \widetilde{u}= e^{-
\gamma_5\alpha/2}u$$ $$ \overline{\widetilde{u}}= \overline{u}
e^{\gamma_5\alpha/2}.
$$
Taking into account that the Dirac bispinors are normalized using
the relation $\overline{u} u =2m$ (see [78]), we have
\be\label{u1u} \overline{\widetilde{u}}\widetilde{u} =\overline{u}
u = 2m. \ee

In conclusion, we note that it is interesting to observe a natural
appearance of the operator ${\cal C}$ in form (100) in the
considered algebraic theory with $\gamma_5$-mass extension (see
[73]). For this purpose we write the modified Dirac equation and
the conjugate equation in $x$ representation, \be \label{psi1}
\left( i\gamma^{\mu}
\partial_{\mu} - (m_1 + m_2 \gamma_5)\right) \widetilde{\psi} = 0 \ee \be \label{psi-}
\overline{\widetilde{\psi}} \left( i\gamma^{\mu}
\partial_{\mu} + (m_1 - m_2 \gamma_5)\right)  = 0. \ee
Let us rewrite them as \be \label{psi} \left( i\gamma^{\mu}
\partial_{\mu} - m \exp{(\alpha\gamma_5))} \right) \widetilde{\psi }= 0 \ee
 \be
\label{psi-1} \overline{\widetilde{\psi}} \left( i\gamma^{\mu}
\overleftarrow{\partial_{\mu}}  + m \exp{(-\alpha\gamma_5)}
\right)  = 0. \ee Now let us multiply the first equation by from
$\overline{\widetilde{\psi}}e^{-\alpha \gamma_5}$ the left, and
the second equation by from $e^{\alpha \gamma_5}\widetilde{\psi}$
the right. Summing the first and the second equations and
performing simple commutations, we obtain the continuity law for
the current density, \be\label{dj}
\partial_{\mu} \left( \overline{\widetilde{\psi}} e^{-\alpha\gamma_5}
\gamma_{\mu} \widetilde{\psi}\right)= 0. \ee In this case, the
current $j_{\mu}$  is defined by the following formula: \be
\label{j} j_\mu = \overline{\widetilde{\psi}}
e^{-\alpha\gamma_5}\gamma_{\mu} \widetilde{\psi}. \ee Taking into
account that $\int_{V_{\infty}} \textrm{div} \overrightarrow{j} dV
= 0, $ we obtain the time conservation of the quantity
$\int_{V_{\infty}} j_0 dV \equiv \int_{V_{\infty}} \rho dV.$ Then,
using the transformations from [73], we have
$$\int_{V_{\infty}} \rho dV = \int \overline{\widetilde{\psi}}
e^{-\alpha\gamma_5} \gamma_0 \widetilde{\psi}\ dV = \int
\overline{\widetilde{\psi}} \eta_0^{-1} {\cal P} \widetilde{\psi}
dV = \int \overline{\widetilde{\psi}} {\cal C} \widetilde{\psi} dV
= 1. $$ Thus, the quantity $\overline{\widetilde{\psi}}{\cal C}
\widetilde{\psi} $ has the meaning of the probability amplitude
and it can be seen that the scalar product contains the operator
${\cal C}= \eta_0^{-1} {\cal P} $. Hence, in this case the
operator ${\cal C}$, coinciding with operator (71) constructed in
strict agreement with the theories [32, 57] appears in a natural
way.

\section{$\gamma_5$-modified Dirac model in a homogeneous  magnetic field}

It is known that exact solutions to Dirac wave equation are the
basis of relativistic quantum mechanics and quantum
electrodynamics of spinor particles in external magnetic fields.
The obtaining, analyzing, and using its exact solutions is an
important aspect of development of this field of science. Just a
few physically important exact solutions to the original Dirac
equations are known. They include, in particular, an electron in a
Coulomb field, a homogeneous magnetic field, and a plane wave
field. Exact solutions to the relativistic wave equation, i.e.,
single-particle wave functions, make it possible to apply the
approach known as the Furry picture which is based on their use.
This very fruitful method of investigation includes the study of
particle interaction with an external field independently of the
field strength value (see [78]). It is interesting to study
non-Hermitian extensions of the Dirac model which represent
alternative formulations of relativistic quantum mechanics and try
to implement the Furry picture method in them [72–74].

Let us consider the homogeneous magnetic field
$\textbf{H}=(0,0,H)$ directed along the axis $x_3$ ($H > 0$). The
electromagnetic field potentials in gauge [78] can be chosen in
the following form:  $A_0=0,\,\,A_1=0,\,\,A_2=H x_1,\,\, A_3=0.$
Let us write the modified Dirac equation in the form
 \be\label{cal P} \left(
\gamma_\mu {\cal P^\mu} -m e^{\vartheta
\gamma_5}\right)\widetilde{\Psi}=0, \ee
 where ${\cal P^\mu} =
i\partial_{\mu} -e A_\mu $ ; $e=-|e|,$ and we use
$\gamma$-$\gamma$-matrices in the standard representation. In the
considered region the integrals of motion ${\cal P}_0,\,{\cal P}_2
$ and ${\cal P}_3$   mutually commute, $[{\cal D},{\cal P}_0]=0$,
$[{\cal D},{\cal P}_2]=0$, $[{\cal D},{\cal P}_3]=0$ where ${\cal
D}=( \gamma_\mu {\cal P}^{\mu} - m e^{\vartheta \gamma_5}) $.

Let us represent the function $\widetilde{\Psi }$ in the form
$$
        \widetilde{ \Psi} = \left(%
\begin{array}{c}
  \psi_1 \\
  \psi_2 \\
  \psi_3 \\
  \psi_4 \\
\end{array}%
\right)e^{-i E t}
$$
and use the Hamiltonian form of the Dirac equations, \be
\label{H7}H\widetilde{\psi} = E\widetilde{\psi}, \ee  where
$$
H=(\overrightarrow{\alpha} \textbf{{\cal P}}) +\beta m_1
+\beta\gamma_5 m_2.
$$
Changing the variables [78]
$$
  \psi_i(x_1,x_2,x_3)= e^{ip_2 x_2+ip_3 x_3}\Phi_i(x_1),
$$
where $i=1,2,3,4,$$i=1,2,3,4,$ we obtain the following system of
equations [73]: \be\label{sist1}(E\mp m_1)\Phi_{1,3}+iR_2
\Phi_{4,2} -(p_3 \mp m_2) \Phi_{3,1} =0; \ee

\be\label{sist}
 (E\mp
m_1)\Phi_{2,4}+iR_1\Phi_{3,1}+(p_3 \pm m_2) \Phi_{4,2}=0. \ee
Here, $R_1=\left[\frac{\partial}{\partial x_1} -(p_2+{e
H})\right],$  $ R_2=\left[\frac{\partial}{\partial x_1} +(p_2+{e
H})\right],$ the upper sign is related to the components of the
wave function with the first index, and the lower sign, to the
components of the wave function with the second index.

It is convenient to use the following dimensionless variable:
\be\rho = \sqrt{\gamma}x_1 +p_2/\sqrt{\gamma},  \ee where
$\gamma=|e| H$, then Eqs. (120), (121) take the form
\be\label{PH1}
 (E\mp m_1)\Phi_{1,3} + i
 \sqrt{\gamma}\left(\frac{d}{d\rho}+\rho\right)\Phi_{4,2}-(p_3 \mp
 m_2)\Phi_{3,1} =0;
\ee \be\label{PH2} (E\mp m_1)\Phi_{2,4} + i
 \sqrt{\gamma}\left(\frac{d}{d\rho}-\rho\right)\Phi_{3,1}+(p_3 \pm
 m_2)\Phi_{4,2} =0.
\ee The general solution to this system can be represented in the
following form:
$$
u_n(\rho)=\left(\frac{\gamma^{1/2}}{2^n n! \pi^{1/2}}
\right)e^{-\rho^2 /2}H_n(\rho),
$$
where $H_n(x)$  are the standard Hermite polynomials,
$$
  {\mathit{H}}_{n}(x)=(-1)^n
e^{x^2/2}\frac{d^n}{dx^n}e^{-x^2/2},
$$
and $n=0,1,2..$.

 Note
that this Hermite functions satisfy the following recurrence
relations: \be\label{u11}
\left(\frac{d}{d\rho}+\rho\right)u_n=(2n)^{1/2}u_{n-1}; \ee
\be\label{u22}
\left(\frac{d}{d\rho}-\rho\right)u_{n-1}=-(2n)^{1/2}u_{n}.\ee It
can be easily seen from (125), (126) that
$$
   \left(\frac{d}{d\rho}-\rho \right)\left(\frac{d}{d\rho}+\rho
   \right)u_n = -2n u_n
$$
 and, therefore (see,
e.g., [78]), \be\label{R1R2}
   R_1 R_2 =-2\gamma n.
\ee
Substituting into (123), (124) the expression
$$
\Phi=\left(%
\begin{array}{c}
  C_1 u_{n-1}(\rho)\\
  iC_2 u_n(\rho)\\
  C_3 u_{n-1}(\rho)\\
  iC_4 u_{n}(\rho)\\
\end{array}%
\right),
$$
it can be obtained that the coefficients $C_i\,(i=1,2,3,4)$ are
determined by the algebraic equations
$$
(E\mp m_1)C_{1,3}-(2\gamma n)^{1/2} C_{4,2} -(p_3 \mp m_2)C_{3,1}
=0;
$$
$$
(E\mp m_1)C_{2,4}-(2\gamma n)^{1/2} C_{3,1}+(p_3 \pm m_2)
C_{4,2}=0.
$$
Let us find the energy spectrum of this non-Hermitian Hamiltonian
by equating to zero the determinant of this system, \be\label{40}
    E=\pm\sqrt{{m_1}^2-{m_2}^2 + 2\gamma n +{p_3}^2},
\ee where $n=0,1,2..$, taking into account that
$m^2={m_1}^2-{m_2}^2$, we can see that this result (see also
(104)) coincides with the eigenvalues of the Hermitian Hamiltonian
describing Landau relativistic levels (see, e.g., [78]).

The coefficients $C_i$ can be determined if we use the third
component of the polarization tensor in the direction of the
magnetic field as the polarization operator [78], \be\label{muH}
          \mu_3=m_1\sigma_3 + \rho_2[\vec{\sigma}\vec{{\cal
          P}}]_3,\ee
where the matrices
$$ \sigma_3= \left(%
\begin{array}{cc}
  I & 0 \\
  0 & -I \\
\end{array}%
\right); \,\,\,\,\,              \rho_2 = \left(%
                                      \begin{array}{cc}
                                        0 & -iI \\
                                     iI & 0 \\
                                         \end{array}%
                                       \right).
$$

It can be easily seen that the bispinor $C$ can be written in the
form

         \be\label{PsiH1} \left(%
\begin{array}{c}
  C_1 \\
  C_2 \\
  C_3 \\
  C_4 \\
\end{array}%
\right)=\frac{1}{2\sqrt{2}}\left(%
\begin{array}{c}
  \mathrm{ch}(\vartheta/2) \Phi_1+\mathrm{sh}(\vartheta/2) \Phi_3 \\
  \mathrm{ch}(\vartheta/2) \Phi_2+\mathrm{sh}(\vartheta/2) \Phi_4 \\
  \mathrm{sh}(\vartheta/2) \Phi_1 +\mathrm{ch}(\vartheta/2) \Phi_3\\
  \mathrm{sh}(\vartheta/2) \Phi_2 +\mathrm{ch}(\vartheta/2) \Phi_4, \\
\end{array}%
\right),\ee  where
$$
\Phi_1=\sqrt{1+\zeta m/p_\bot}\sin(\pi/4+\lambda/2)
$$
$$
\Phi_2=\zeta\sqrt{1-\zeta m/p_\bot}\sin(\pi/4-\lambda/2)
$$
$$
\Phi_3=\zeta\sqrt{1+\zeta m/p_\bot}\sin(\pi/4-\lambda/2)
$$
$$
\Phi_4=\sqrt{1-\zeta m/p_\bot}\sin(\pi/4+\lambda/2).
$$
Here, $\mu_3 \psi =\zeta k\psi$, $k=\sqrt{{p_\bot}^2 + m^2}$ and
$\zeta=\pm 1 $,  which corresponds to the following fermion spin
orientation: along $(+1)$  and opposite $(-1)$ to the magnetic
field, and the parameter $ \lambda$ satisfies the relation
$\cos{\lambda}=p_3/E.$ The functions $\mathrm{sh}(\vartheta/2)$
and $\mathrm{ch}(\vartheta/2)$  are defined by the relations
\be\label{chsh} \mathrm{ch}\vartheta = m_1/m;\,\,\,\,
\mathrm{sh}\vartheta = m_2/m. \ee 

It should be pointed out that the use of the exact solution to the
Dirac equation in a magnetic field has recently yielded a number
of interesting experimental results. In particular, in [79, 80]
exact agreement between this relativistic model and different
combinations of Jaynes–Cummings (JC) [79] and anti-Jaynes–Cummings
(AJC) [80] interactions was found. These nontrivial facts imply
important results which are widely used in quantum optics.

\section{Modified non-Hermitian Dirac-Pauli model in a magnetic field}

In this section, we consider the description of motion of Dirac
particles with magnetic momentum other than the Bohr magneton [3,
4]. It was shown by Schwinger [81] that the Dirac equation for
particles in an external electromagnetic field $A^{ext}$  with
account of radiative corrections can be represented as
\be\label{A} \left(\hat{{\cal P}}\gamma -m\right)\Psi(x)-\int{\cal
M}(x,y|A^{ext})\Psi(y)dy=0, \ee where ${\cal M}(x,y|A^{ext})$  is
the mass operator of fermions in an external field. The modified
equation (see, e.g. [78]) can be obtained from Eq. (132) by
expanding the mass operator in $ eA^{ext}$ series to the first
order with respect to the field. This equation possesses
relativistic covariance and agrees with the phenomenological Pauli
equation obtained in his early papers.

Let us consider the model of massive fermions with
$\gamma_5$-mass extension $m\rightarrow m_1+\gamma_5 m_2$ , taking
into account the interaction of their charges and anomalous
magnetic moments with the electromagnetic field $F_{\mu\nu}$:

\be\label{Delta} \left( \gamma^\mu {\cal P}_\mu -
 m_1 -\gamma_5 m_2 -\frac{\Delta\mu}{2}\sigma^{\mu \nu}F_{\mu\nu}\right)\widetilde{\Psi}(x)=0,\ee
where  $\Delta\mu = (\mu-\mu_0)= \mu_0(g-2)/2$. Here, $\mu$  is
the magnetic moment of the fermion, $g$ is the gyromagnetic factor
of the fermion, $\mu_0=|e|/2m$ is the Bohr magneton, and and
$\sigma^{\mu\nu}=i/2(\gamma^\mu \gamma^\nu-\gamma^\nu
\gamma^\mu)$.  Thus, the phenomenological constant $\Delta\mu $
introduced by Pauli is a part of the equation and can be
interpreted from the point of view of the quantum field theory.

The Hamiltonian form of Eq. (133) for fermions in a homogeneous
magnetic field is [3, 4]: \be i\frac{\partial}{\partial t}
\widetilde{\Psi}(r,t)=H_{\Delta \mu}\widetilde{\Psi}(r,t),\ee
where \be\label{Delta1} H_{\Delta\mu} = \vec{\alpha}\vec{{\cal P}}
+ \beta(m_1 + \gamma_5 m_2) +
\Delta\mu\beta(\vec{\sigma}\textbf{H}).\ee In particular, taking
into account the quantum electrodynamic contribution into the
anomalous magnetic moment of the electron to $e^2$, we have
$\Delta\mu=\frac{\alpha}{2\pi}\mu_0 $,  where $\alpha = e^2
=1/137$ is the fine structure constant, and it is still assumed
that the external field potential satisfies the free Maxwell
equation.

It should be noted that here the operator of fermion spin
projection onto the direction of its motion $\overrightarrow{
\sigma} \overrightarrow{{\cal P }} $  does not commute with
Hamiltonian (135) and therefore, is not an integral of motion. The
operator commuting with the Hamiltonian is $\mu_3$ (see (129)).
Thus, the wave function $\widetilde{ \psi }$ is the eigenfunction
of polarization operator (129) and Hamiltonian (135). Therefore,
we have \be\label{Pi}
\mu_3\psi = \zeta k\psi, \,\,\, \mu_3=\left(%
\begin{array}{cccc}
  m_1 & 0 & 0 & {\cal P}_1-i{\cal P}_2 \\
  0 & -m_1 & -{\cal P}_1-i{\cal P}_2 & 0 \\
  0 & -{\cal P}_1+i{\cal P}_2 & m_1 & 0 \\
  {\cal P}_1+i{\cal P}_2 & 0 & 0 & -m_1 \\
\end{array}%
\right), \ee where $\zeta=\pm 1$  characterizes the fermion spin
projection onto the magnetic field direction, and

$$H_{\Delta\mu}\widetilde{\psi}=E\widetilde{\psi},$$
  \be\label{Hmu} H_{\Delta\mu}=\left(%
\begin{array}{cccc}
  m_1+H\Delta\mu & 0 & {\cal P}_3 -m_2& {\cal P}_1-i{\cal P}_2 \\
  0 & m_1-H\Delta\mu & {\cal P}_1+i{\cal P}_2  & -m_2-{\cal P}_3\\
  m_2+{\cal P}_3 & {\cal P}_1-i{\cal P}_2 & -m_1-H\Delta\mu & 0 \\
  {\cal P}_1+i{\cal P}_2 & m_2-{\cal P}_3 & 0 & H\Delta\mu-m_1 \\
\end{array}%
\right). \ee

These calculations are similar to the calculations given in detail
in the previous section. As a result, with the help of the
modified Dirac–Pauli equation we can also find \emph{the exact
values of the energy spectrum} in the case of non-Hermitian
Hamiltonian (135), (137) (see [3, 74]),
 \be\label{E61} E(\zeta,p_3,2\gamma
n,H)=\pm\sqrt{{p_3}^2-{m_2}^2+\left[\sqrt{{m_1}^2+2\gamma
n}+\zeta\Delta\mu H \right]^2}, \ee and the eigenvalues of the
polarization operator $\mu_3$
 \be k=\sqrt{{m_1}^2 +2\gamma n}.
\ee

It should be noted that formula (138) is satisfied not only for
charged fermions but also for neutral particles with anomalous
magnetic moment. In this case it is necessary to replace the value
of the quantized transverse momentum of a charged particle in a
magnetic field by the regular value, $2\gamma n\rightarrow
{p_1}^2+{p_2}^2={p_\bot}^2$. As a result, we obtain [4, 73]:
 \be\label{E62} E(\zeta,p_3,p_\bot,H)=\pm\sqrt{{p_3}^2-{m_2}^2+\left[\sqrt{{m_1}^2+
 {p_\bot}^2}+\zeta\Delta\mu H \right]^2}. \ee

It can be seen from (140) that in the region of unbroken ${\cal
PT}$-symmetry the energy spectrum is real if the particle spin is
directed along the magnetic field, $\xi = +1$. It can be easily
noted, however, that in the case $\zeta=-1$ (the fermion spin is
directed opposite to the magnetic field) in the linear
approximation with respect to the magnetic field strength, real
values of the energy spectrum can be obtained only if the magnetic
field strength is bounded by the following value: \be \label{Hmax}
H\leq \frac{{p_0}^2}{2\Delta\mu\,k}, \ee where $p_0 =\sqrt{m^2
+{p_\bot}^2 +{p_3}^2}$   is the regular Hermitian particle energy.

Therefore, expression (140) yields for $\zeta= -1$ and
$p_\bot=p_3=0$, which corresponds to a fermion at rest, that
$H\leq H_{max}$.  In the linear approximation with respect to the
magnetic field strength we can also find [3, 74] \be\label{33}
H_{max}=\frac{m^2}{2 m_1\Delta\mu}.\ee

It can be easily obtained from (142) that the direct consequence
of Eq. (138) is that the energy eigenvalues for $H\leq H_{max}$
are real. We can formulate a new condition of unbroken ${\cal
PT}$-symmetry for the model of fermions with the non-Hermitian
$\gamma_5$-mass extension and nonzero anomalous magnetic moment in
a high-intensity external magnetic field using (142). This
condition replaces condition (76) and can be written as
\be\label{7} {m_1}^2 -{m_2}^2 \geq 2 m_1 \Delta \mu H.\ee

Thus, it can be seen that there exists the maximum magnetic field
value $H_{max}$ such that if this value is exceeded for fermions
at rest with $\zeta = -1$ it results in the loss of the real
character of the energy spectrum.

It can be seen from (140) [4] that in the region of unbroken
${\cal PT}$-symmetry all energy levels are real if the particle
spin is directed along the magnetic field, $\xi = +1$. If the spin
is directed opposite to the magnetic field, $\xi = -1$, the energy
of the fermion ground state $n=0$ has an imaginary part, as well
as other lower energy levels. Figures 4 and 5 (see also [74]) show
the energy values as functions of the parameter  $x= m/M$ (see
Fig. 4 for $\xi = +1$ and Fig. 5 for $\xi = -1$).

\begin{figure}[h]
\vspace{-0.2cm} \smallskip
\includegraphics[angle=0, scale=0.5]{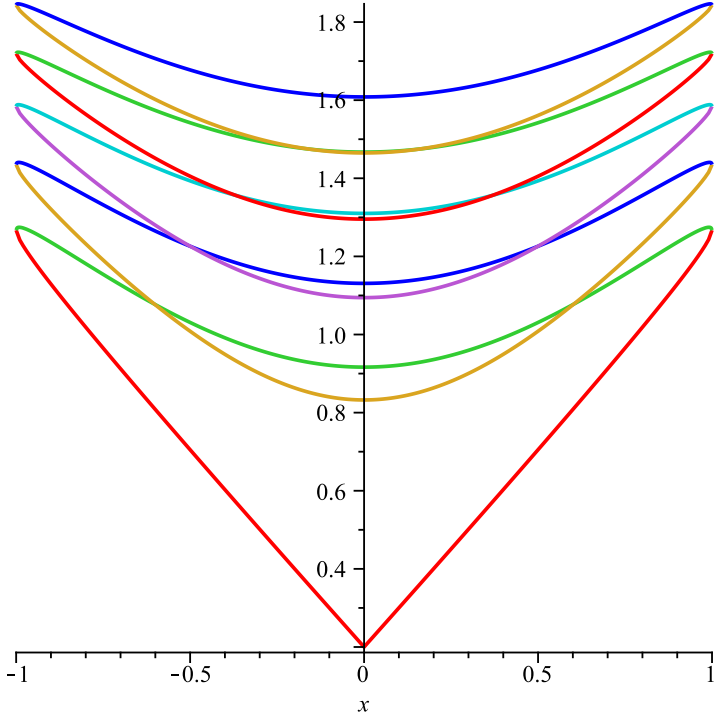}
\caption{Dependence of $E(+1, 0, 0.4n, 0.2)$ on parameter $x=m/M$
for the case $n=0, 1, 2, 3, 4$ and $\Delta\mu H = 0.2.$ }
\vspace{-0.1cm}\label{f6}
\end{figure}

\begin{figure}[h]
\vspace{-0.2cm} \smallskip
\includegraphics[angle=0, scale=0.5]{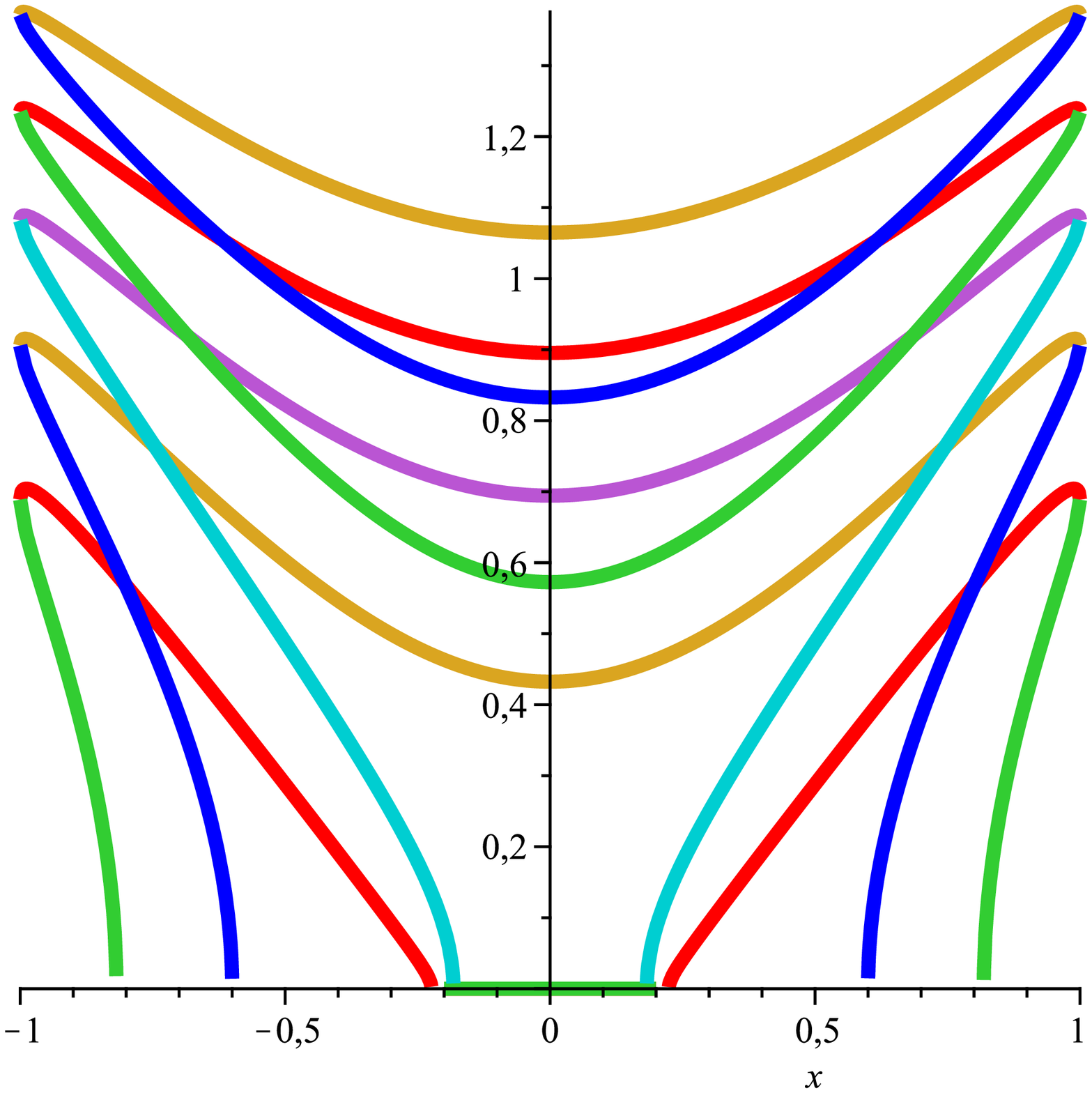}
\caption{Dependence of $E(-1, 0, 0.4n, 0.2)$ on parameter $x=m/M$
for the case $n=0, 1, 2, 3, 4$ and $\Delta\mu H = 0.2.$ }
\vspace{-0.1cm}\label{f7}
\end{figure}

It can be easily seen that in the case $ \Delta\mu =0$  regular
expression for energy levels of a charged particle in an external
field (128) (Landau levels) can be obtained from (138). It should
be underlined that the expression similar to (138) can be obtained
in the framework of the standard Dirac-Pauli approach, assuming
$m_2=0$ and $m_1=m$  (Hermitian limit), \be\label{DPA}
E(\zeta,p_3,2\gamma
n,H)=\pm\sqrt{{p_3}^2+\left[\sqrt{{m}^2+2\gamma n}+\zeta\Delta\mu
H \right]^2}.
 \ee

Note that the result similar to (144) was obtained earlier in [82]
using the regular Hermitian approach to solution of the
Dirac-Pauli equation. Direct comparison of modified formula (144)
in the Hermitian limit (see [3, 4]) with the result obtained in
[82] demonstrates their agreement. It can be easily seen that Eq.
(138) contains the separate dependence on the parameters $m_1$ and
$m_2$, rather than one parameter united in the \emph{physical mass
(of the particle)} $m=\sqrt{{m_1}^2-{m_2}^2}$, which differs
essentially from the examples considered in the previous sections.

Thus, unlike (104) and (128), in this case the calculation of
interaction of the fermion anomalous magnetic moment with the
magnetic field makes it possible to put forward the question of
the possibility of experimental observation of effects of
$\gamma_5$-extension of the fermion mass.

Note that if it is assumed that $m_2=0$ and therefore $m_1 =m$, we
obtain, as it was already indicated earlier, the Hermitian limit.
Taking into account expressions (81) and (82), however, it can be
obtained that energy spectrum (138) is expressed in terms of the
fermion mass $m$ and the maximal mass $M$. Thus, taking into
account that the interaction of anomalous magnetic moment with the
magnetic field eliminates degeneration with respect to the spin
variable, we can obtain the energy of the ground state
($\zeta=-1$) in the form (see [3, 4])
 \be\label{E1}
E(-1,0,0,H,x)=m\sqrt{-\left({\frac{1\mp\sqrt{1-x^2}}{x}}\right)^2+
\left(\frac{\sqrt{2}\sqrt{1\mp\sqrt{1-x^2}}}{x}-\frac{\Delta\mu
H}{m} \right)^2}, \ee where $x=m/M$, the upper sign corresponds to
ordinary particles, and the lower sign corresponds to their
“exotic” partners.

\begin{figure}[h]
\vspace{-0.2cm} \smallskip
\includegraphics[angle=0, scale=0.5]{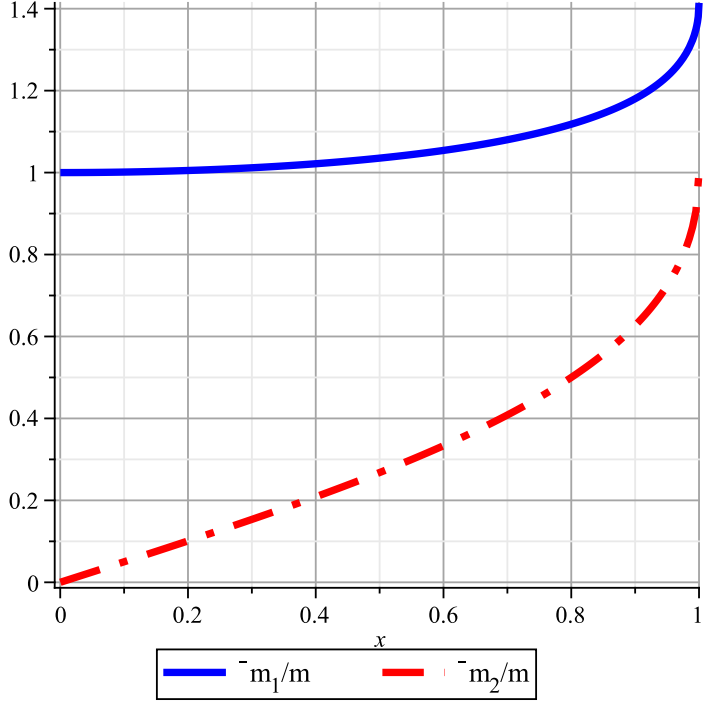}
\caption{Dependence  of $m^{-}_1/m,\, m^{-}_2/m $ on parameter
$x=m/M.$} \vspace{-0.1cm}\label{f4}
\end{figure}

\begin{figure}[h]
\vspace{-0.2cm} \smallskip
\includegraphics[angle=0, scale=0.5]{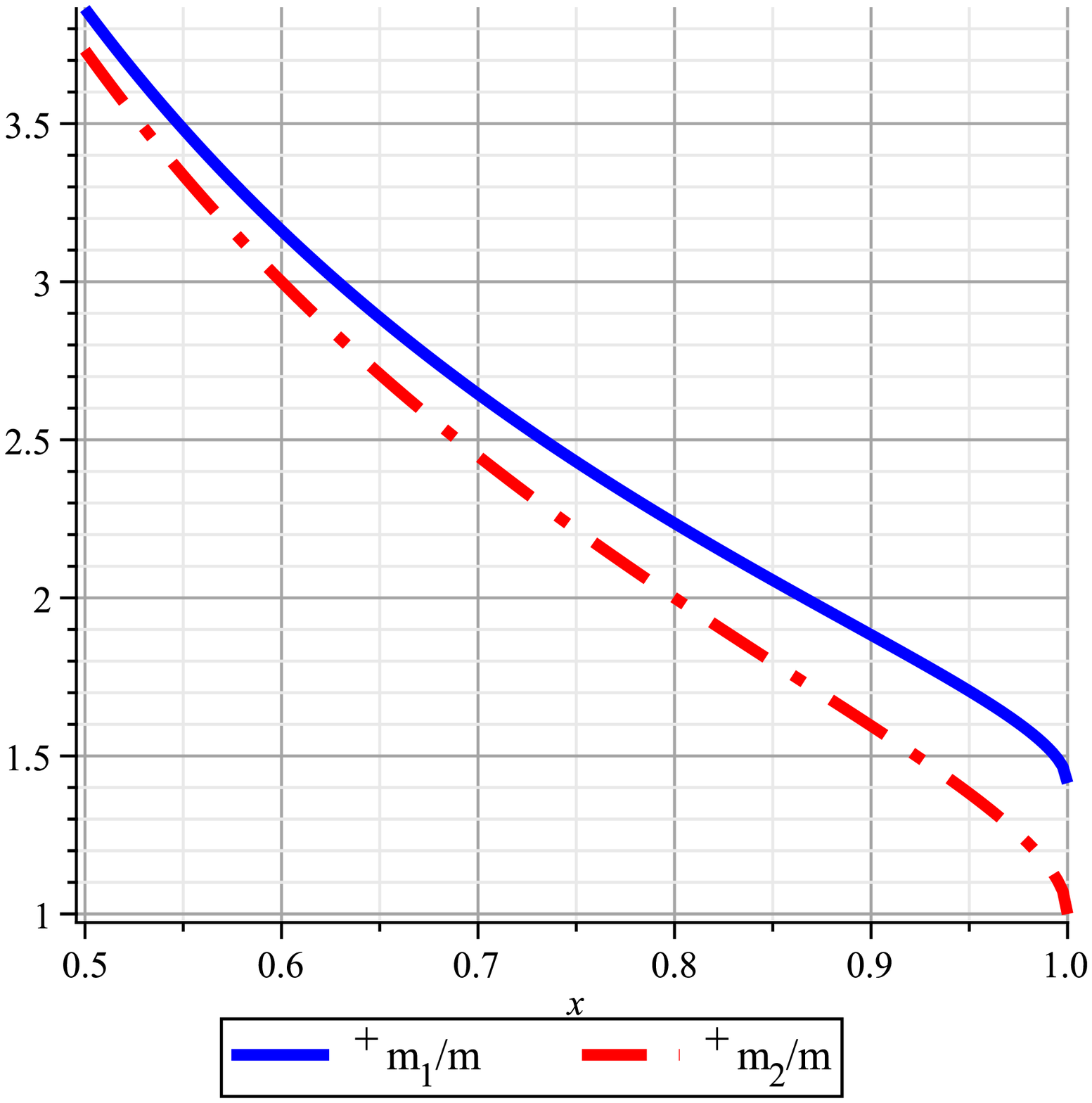}
\caption{ Dependence of parameters $m^{+}_1/m,\, m^{+}_2/m $ on
value $x=m/M.$} \vspace{-0.1cm}\label{f5}
\end{figure}

Let us now pass over to more detailed examination of the ground
state energy of fermions in an external field. Using the
calculations given above (see (138)), the dependence of the energy
of an ordinary fermion with the small mass $x =m/M \ll 1$ on the
magnetic field strength $H$ can be obtained, \be\label{E1}
E(-1,0,0,H,x)=
m\sqrt{1-\frac{\Delta\mu}{\mu_0}\frac{H}{H_c}(1+x^2/8 + 7 x^4/128)
+ \left(\frac{\Delta\mu H}{m}\right)^2},  \ee where $H_c = m^2/e =
4.41\cdot10^{13}$ G is the quantizing magnetic field for an
electron [78].

At the same time, for the case of "exotic" particles in the same
limit $x\ll1$ we have the result quite different from (146) (see
[3, 4]), \be\label{E2} E(-1,0,0,H,x)= m
\sqrt{1-\frac{\Delta\mu}{\mu_0}\frac{2
 H}{x H_c}+ \left(\frac{\Delta\mu H}{m}\right)^2}. \ee
It can be seen from (147) that in this case the field corrections
increase considerably, as $1/x=M/m \gg 1$. In the limit $ m
\rightarrow M$   the results for ordinary and exotic particles
agree. Thus, combining these results we can write the following:
\be\label{E3}
  E(-1,0,0,H,x)= m\sqrt{1-\frac{\Delta\mu}{\mu_0}\frac{\sqrt{2}
 H}{x H_c}+ \left(\frac{\Delta\mu H}{m}\right)^2}.
 \ee

It can also be seen [73], that the change of parameters $m^{\mp}_1
$ and $m^{\mp}_2$  takes place in such a way that at the point $
x=1$ ($m=M$) the branches of the plots for the mass parameters of
ordinary and exotic particles inter sect. It can be easily seen in
Figs. 6 and 7 showing ${m}^{-}_{1, 2}/m$  and $m^{+}_{1, 2}/m$ as
functions of $x=m/M$.

Since Eqs. (133) and formulas (138), (145), and (147) following
from these equations are valid for practically any magnetic field
intensities, it can be easily seen that for \be H\sim \frac{\mu_0
}{\Delta\mu}\frac{m}{m_1} H_c\ee we obtain $E_0\sim 0$.  Thus, in
a high-intensity magnetic field, consideration of the vacuum
magnetic moment may result in an essential change of boundary of
the energy spectrum between fermion and anti-fermion states. It
should be pointed out that a considerable growth of this
correction is connected with the possible contribution of the
so-called exotic particles [3].

Let us consider the neutrino as an important example of applying
the obtained expressions for the energy. Namely, in the case of
ultracold polarized regular electron neutrinos, we obtain in the
linear approximation with respect to the field (see (140))

\be\label{E34} E(-1,0,0,H,m_{\nu_e}/M \ll 1 )= m_{\nu_e}
\sqrt{1-\frac{\mu_{\nu_e}}{\mu_0}\frac{
 H}{ H_c}}.
 \ee

In the case of exotic electron neutrinos, however, the situation
may change essentially,

 \be\label{E3}
E(-1,0,0,H,m_{\nu_e}/M \ll 1)= m_{\nu_e}
\sqrt{1-\frac{\mu_{\nu_e}}{\mu_0}\frac{2 M
 H}{m_{\nu_e} H_c}}.
 \ee

It is well known [83, 84] that in the minimally extended SM
one-loop radiative corrections contributing to the neutrino
magnetic moment are proportional to the neutrino mass,
\be\label{mu1}
  \mu_{\nu_e}=\frac{3}{8\sqrt{2}\pi^2}|e| G_F
  m_{\nu_e}=\left(3\cdot10^{-19}\right)\mu_0\left(\frac{m_{\nu_e}}{1
  eV}\right),
\ee where $ G_F$ is the Fermi interaction constant, and $\mu_0$ is
the Bohr magneton. The discussion of the problem of measurement of
the neutrino mass (active or sterile components) yields that for
the active neutrino of the model we have $\sum m_\nu =0.320 eV$,
while for the sterile neutrino, $\sum m_\nu =0.06 eV$ [85].

Thus, the change of boundary of the region of unbroken ${\cal
P}{\cal T}$ - symmetry due to the shift of the state with the
lowest energy in the magnetic field can be estimated. Indeed, Eqs.
(150) and (151) can be used to define the regions of ${\cal
P}{\cal T}$ - symmetry preservation.

Let us consider the following neutrino parameters: the mass of the
electron neutrino $m_{\nu_e} = 1 eV$ and magnetic moment (152). If
it is assumed that the mass and magnetic moment of exotic
neutrinos are identical to the parameters of ordinary neutrinos,
the estimates for the boundaries of the region of unbroken ${\cal
P}{\cal T}$ symmetry can be obtained for (150) in the form [3]
\be\label{E4}
 {H^{max}}_{\nu_e(ordinary)}  =  \frac{\mu_0}{\mu_{\nu_e}} H_c.\ee
In case (148), however, the situation may change radically,
\be\label{E41}
 {H^{max}}_{\nu_e(exotic)}  =  \frac{\mu_0}{\mu_{\nu_e}} \frac{m_{\nu_e}}{2 M} H_c.\ee
It may turn out that critical magnetic field value (154) can be
reached in the ground-based experiments [3, 4], unlike (153),
where the experimentally determined field corrections are
extremely small.

Note also that high intensity magnetic fields exist near and
inside some cosmological objects. Namely, magnetic fields with a
strength of order of $10^{12}\div10^{13}$ G are observed near
pulsars. Such recently discovered objects as soft gamma repeaters
and anomalous X ray pulsars can also be attributed to the objects
of interest. Magnetic rotating models called magnetars are
proposed for such objects. It was demonstrated that the strength
of magnetic fields for these objects can reach up to  $10^{15}$ G.
It is extremely important that the fraction of magnetars among
neutron stars can reach 10\% [86, 87]. In this relation, we note
that processes with participation of ordinary neutrinos and
especially their possible “exotic partners” in the presence of
such strong magnetic fields can essentially influence the
processes defining the development of astrophysical objects.

Thus, finalizing the abovesaid, it should be noted that the main
result obtained in the course of algebraic construction of the
fermion model with $\gamma_5$  mass extension is that the new mass
(energy) scale determined by the parameter $m_{max}\equiv
M={m_1}^2/2m_2$ was introduced. This value on the mass scale is
the point of transition from ordinary to exotic particles.
Actually, the algebraic approach contains the same description of
exotic particles as the geometric models with a fundamental mass.

It should be noted that, although energy spectra of fermions in
some cases coincide with the spectra of the corresponding
Hermitian Hamiltonians  $ H_0 $, we found examples in which the
fermion energy depends explicitly on non-Hermitian parameters: for
instance, the study of interaction of fermion anomalous magnetic
moment with a magnetic field. In this case, we obtained the exact
solution to the fermion energy spectrum (see (138, (140)).

We do not know whether the upper limit of the mass spectrum of
elementary particles is equal to the Planck mass [5], but
experimental study of this variant of the theory at high energies
can hardly be considered at present. The alternative modern
precision laboratory measurements at low energies in
high-intensity magnetic fields, in principle, make it possible to
achieve the required values for exotic particles if they exist.
Formulas (151)–(154) (see [3, 4]) prove the existence of a
\emph{maximal mass} and reality of the so-called \emph{exotic
particles}, since the latter are inseparably connected with the
constraint for the mass spectrum of elementary particles.

\section{Conclusions}

The consistent application of Markov’s concept assuming the
existence of a stable object with the Planck mass and stating the
existence of a finite upper limit $M$  for the mass spectrum of
elementary particles lead Kadyshevsky to the development of a
modified version of the local quantum field theory based on the
geometric approach. In this model, the parameter $M$  was not only
the limiting admissible mass of a particle but was also manifested
as a new universal energy scale. Along with the abovesaid,
Kadyshevsky also investigated a number of problems connected with
the possible existence of a fundamental length constant which was
marked by P.A. Dirac as one of the most important problems of
modern physics. In this context, the “fundamental length” served
as the parameter complementary to the “fundamental mass”, thus
being extremely important for establishing the bound-aries of
applicability of modern physical theories.

Thus, practically all publications by Kadyshevsky clearly
demonstrate his devotion to geometric scenarios of QFT development
and extension. He assumed that theories based on the geometric
principle had real chances to be logically consistent. Einstein’s
heuristic formula

\begin{center} \textbf{“Experiment = Geometry + Physics”}
\end{center}
 more than once appeared in our
discussions, and Kadyshevsky was sure that this formula was true.
The power of this approach has been proved in our joint studies on
development of the local QFT based on de Sitter geometry
containing a hypothetic universal parameter $M$ [15, 17]. It seems
absolutely justified to rank Kadyshevsky’s works together with
those of prominent physicists who studied the problems of
existence of fundamental constraints with the dimension of mass
and length [88].

The introduction of the constraint on the mass spectrum $m\leq M$
based on the geometric approach to development of the modified QFT
results in the appearance of non-Hermitian (pseudo-Hermitian)
$\cal PT$ symmetric Hamiltonians in the fermion sector of the
model with the same region of unbroken $\cal PT$ symmetry, $m\leq
M$. The extensive development of the theory of non-Hermitian
quantum models made it possible to overcome the difficulties in
the theory due to its non-Hermitian character.

The synthesis of the geometric theory with a maximal mass and the
algebraic theory with $\gamma_5$ extension for the mass of a Dirac
particle turned out extremely fruitful for both theories. In
particular, introducing the parameter of maximal mass and
postulating its equality to the fundamental mass $M$ makes it
possible to impart physical meaning to the algebraic theory and
describe the whole spectrum of particles known in SM with its
help. At the same time, the application of methods developed in
the algebraic model actually makes a breakthrough in solving the
problems connected with this theory. We mean both the mathematical
problems brought forward by the non-Hermitian character of the
considered operators and the problems of formulating experiments
on verification of the theory and the search of the value of $M$
(see [3, 4]).

In particular, it turns out that an experiment on verification of
a theory with a maximal mass should not necessarily be performed
at superhigh energies, as it was assumed earlier, but can be
performed in the low energy region. Here, we mean the interaction
of anomalous magnetic moment of fermions with a high-intensity
magnetic field, namely, the behavior of ultra-cold polarized
neutrinos in an external magnetic field. The “moment of truth” in
this experiment is the verification of existence of the so-called
“exotic” neutrinos [4] whose description does not proceed to the
Dirac limit $M\rightarrow \infty$ (or the “planar limit” in the
geometric interpretation).

Thus, an experiment on verification of the theory with a maximal
mass and elucidation of whether it is necessary to go beyond the
framework of the standard quantum field theory, in principle, can
be performed in near future. There is an assumption that “exotic”
particles are a part of “dark matter” which is known to make a
large fraction of energy density of the Universe and cannot be
described in the framework of SM. This means that the development
of a theory with a limited mass as a modification and extension of
modern QFT can be an important stimulus in solving this problem.

\end{document}